\documentclass[prb,twocolumn,amsmath,amssymb,superscriptaddress]{revtex4-2}

\usepackage[pdftex]{graphics}
\usepackage{graphicx}
\usepackage{times}
\usepackage{color}
\usepackage{enumerate}
\usepackage{ulem}

\begin{document}
	
\title{Spin-orbit coupling and crystal-field splitting in Ti-doped Ca$_2$RuO$_4$ studied by ellipsometry}
\author{I. Vergara}
\author{M. Magnaterra}
\affiliation{Institute of Physics II, University of Cologne, 50937 Cologne, Germany }
\author{J. Attig}
\affiliation{Institute for Theoretical Physics, University of Cologne, 50937 Cologne, Germany}
\author{S. Kunkem\"{o}ller}
\author{D.I. Khomskii}
\author{M. Braden}
\affiliation{Institute of Physics II, University of Cologne, 50937 Cologne, Germany }
\author{M.~Hermanns}
\affiliation{Department of Physics, Stockholm University, AlbaNova University Center, SE-106 91 Stockholm, Sweden}
\affiliation{Nordita, KTH Royal Institute of Technology and Stockholm University, SE-106 91 Stockholm, Sweden}
\author{M. Gr\"{u}ninger}
\affiliation{Institute of Physics II, University of Cologne, 50937 Cologne, Germany }

\date{March 24, 2022}

\begin{abstract}
In Ca$_2$RuO$_4$, the competition of spin-orbit coupling $\zeta$ and tetragonal crystal field 
splitting $\Delta_{\rm CF}$ has been discussed controversially for many years.  
The orbital occupation depends on $\Delta_{\rm CF}/\zeta$, which allows us to address this ratio via 
the optical spectral weights of the lowest intersite Mott-Hubbard excitations.   
We study the optical conductivity of Ca$_2$Ru$_{0.99}$Ti$_{0.01}$O$_4$ in the range of 
0.75 -- 5\,eV by ellipsometry, using the large single crystals that can be grown 
for small Ti concentrations. 
Based on a local multiplet calculation, our analysis results in $2.4 \leq \Delta_{\rm CF}/\zeta \lesssim 4$ at 15\,K.\@ 
The dominant crystal field yields a ground state 
close to $xy$ orbital order but spin-orbit coupling is essential for a quantitative description of the properties. 
Furthermore, we observe a pronounced decrease of $\Delta_{\rm CF}$ with increasing temperature, 
as expected based on the reduction of octahedral distortions. 
\end{abstract}

\maketitle

\section{Introduction}

In $5d$ transition-metal compounds, the interplay of strong spin-orbit coupling and electronic correlations 
gives rise to novel electronic phases \cite{WitczakKrempa14,Rau16,Schaffer16,Takayama21,Khomskii21}. 
In materials with a partially filled $4d$ shell, spin-orbit coupling is much smaller, which may shift the balance 
in the competition between spin-orbit coupling, crystal field, electronic correlations, and exchange interactions \cite{Streltsov20}. 
We focus on the layered $4d^4$ compound Ca$_2$RuO$_4$ which features a temperature-driven metal-insulator 
transition at $T_{\rm MI}$\,=\,357\,K \cite{Nakatsuji97,Alexander99} and antiferromagnetic (AF) order 
below the N\'eel temperature $T_N$\,=\,110\,K \cite{Braden98}.
Remarkably, it can be driven into a conducting non-equilibrium phase by small electric fields or 
currents \cite{Nakamura13,Okazaki13,ZhangPRX19,Bertinshaw19,Jenni20}. 
The phase diagram of Ca$_{2-x}$Sr$_x$RuO$_4$ is rich \cite{Nakatsuji00a,Nakatsuji00b,Nakatsuji03,Nakatsuji04,Carlo12,Ricco18} 
even though the formal valence of Ru$^{4+}$ is independent of $x$, pointing towards a prominent role of the $x$-dependent tilt and rotation angles of the RuO$_6$ octahedra \cite{Friedt01}. 
The most controversially discussed issue in Ca$_2$RuO$_4$ is the relative size of spin-orbit coupling $\zeta$\,=\,$2\lambda$ and tetragonal crystal field splitting $\Delta_{\rm CF}$, which affects for instance the 
characters of the magnetic moments and of the magnetic excitations 
\cite{Khaliullin13,Akbari14,Kunkemoeller15,Kunkemoeller17,Jain17,Souliou17,Zhang17,Zhang20,Sarte20,Mohapatra20,Feldmaier20,Strobel21}.

\begin{figure}[t]
\includegraphics[width=0.9\columnwidth,clip]{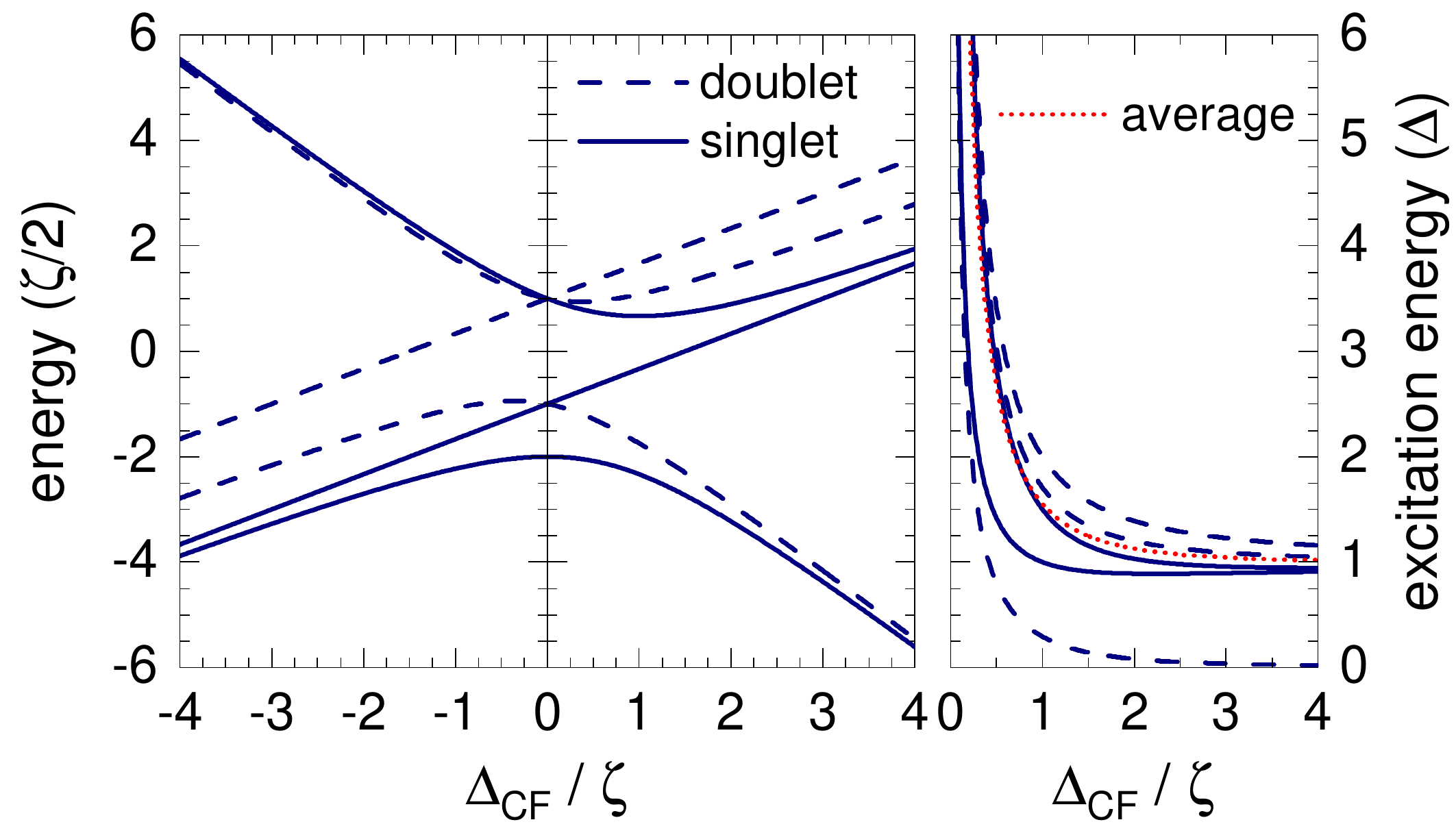}
	\caption{Left: Energies of the nine low-energy states for a single $t_{2g}^4$ site 
	with Coulomb interactions, spin-orbit coupling $\zeta$, and tetragonal crystal-field splitting $\Delta_{\rm CF}$. 
	For $\Delta_{\rm CF}$\,=\,0, the states group into $J$\,=\,0, 1, and 2. 
	For $\Delta_{\rm CF} \gg \zeta$, the three lowest states extrapolate to the $S$\,=\,1 triplet with double occupancy of the $xy$ orbital.
	Right: Excitation energies. Red dotted line shows the average of the upper four energies. 
}
	\label{fig:energiesLS}
\end{figure}

In the limit $\zeta$\,=\,0 with $\Delta_{\rm CF}$\,$>$\,0, the tetragonal crystal field favors a spin $S$\,=\,1 state with double occupancy of the $xy$ orbital, so-called $xy$ orbital order, which has been found in several first-principles studies \cite{Jung03,Fang04,Gorelov10}. 
In the $S$\,=\,1 picture, spin-orbit coupling is treated as a perturbation. Early evidence for its relevance was provided by an x-ray absorption study \cite{Mizokawa01} of the orbital occupation.  
In this picture, 
the highly anomalous spin-wave dispersion can be well described by large single-ion anisotropy terms \cite{Kunkemoeller15,Zhang17,Zhang20}.
In the opposite limit $\Delta_{\rm CF}$\,=\,0 and finite $\zeta$, the local $t_{2g}^4$ configuration adopts 
a $J$\,=\,0 ground state with equal population of the three $t_{2g}$ orbitals. 
In fact, the ground state of a single-site model is a singlet for \textit{any} value of $\Delta_{\rm CF}/\zeta$, 
see Fig.\ \ref{fig:energiesLS}. 
Such a non-magnetic ground state is realized in $5d^4$ iridates with dominant spin-orbit coupling \cite{Yuan17,Fuchs18}. 
In this case, van-Vleck-type excitonic magnetism, or singlet magnetism \cite{KhomskiiBook}, 
may arise if exchange interactions are strong enough to allow for 
condensation of the dispersive lowest excited state \cite{Khaliullin13,Akbari14,Meetei15}. 
In this picture, spin-orbit coupling allows for a longitudinal magnetic mode that corresponds to amplitude fluctuations 
equivalent to a Higgs mode \cite{Khaliullin13,Jain17,Souliou17,Sarte20}.

Excitonic magnetism initially was proposed for the cubic case with $\Delta_{\rm CF}$\,=\,0 \cite{Khaliullin13} 
but its realization is facilitated by reducing the energy of the lowest excited state with increasing 
$\Delta_{\rm CF}/\zeta$ \cite{Akbari14,Feldmaier20}, see Fig.\ \ref{fig:energiesLS}. 
For Ca$_2$RuO$_4$, the two controversially discussed scenarios hence are aiming to describe two sides 
of the same coin. 
On the one hand, LDA+DMFT finds that dominant $xy$ orbital order prevails in the presence of spin-orbit coupling 
and the spin-wave dispersion can be described by interacting local moments 
in an $S$\,=\,1 low-energy model with pronounced single-ion anisotropy
\cite{Zhang17,Zhang20,Kunkemoeller15}.  
Concerning the electronic structure as observed in ARPES, an extension of the anisotropic 
$S$\,=\,1 picture to a $t$-$J$-like model describes the less dispersive bands with $xz/yz$ character \cite{Sutter17,Klosinski20}.
On the other hand, any calculation starting from a single site and local multiplets 
with Coulomb interactions has to deal with the $J$\,=\,0 character of the lowest local state. 
The central quantity of the single-site model is the ratio $\Delta_{\rm CF}/\zeta$ which determines 
the wavefunctions of the local low-energy states. 
From theory, $\Delta_{\rm CF}/\zeta$\,$\approx$\,3 was obtained in 
LDA+DMFT \cite{Zhang20}.
The analysis of on-site $dd$ excitations observed at 12\,K by resonant inelastic x-ray scattering (RIXS) 
at the Ru $L$ edge \cite{Gretarsson19} yields $\zeta$\,=\,0.13\,eV and $\Delta_{\rm CF}/\zeta$\,=\,2.  
In this parameter range, the local $J$\,=\,0 ground state shows dominant occupation of the $xy$ orbital 
and the energy scale of the lowest excitations is small, $(\zeta/2)^2/\Delta_{\rm CF}$.

We demonstrate that optical measurements provide a sensitive tool to determine 
$\Delta_{\rm CF}/\zeta$ in $4d$ compounds. We employ ellipsometry which profits from the availability 
of large crystals since the oblique angle of incidence reduces the effective sample size. 
In-depth studies of Ca$_2$RuO$_4$ were hampered for a long time by the small size of the 
available single crystals. Upon cooling down after crystal growth, the samples typically pulverize 
at the metal-insulator transition at $T_{\rm MI}$\,=\,357\,K due to large accompanying jumps 
of the lattice parameters \cite{Friedt01}. Substituting 1\,\% of the Ru ions by Ti ions 
broadens the phase transition, keeping the crystals intact, while $T_{\rm MI}$ and structural 
and magnetic properties such as the N\'eel temperature $T_N$ are hardly affected \cite{Kunkemoeller17}.

In Ca$_2$Ru$_{0.99}$Ti$_{0.01}$O$_4$, we address $\Delta_{\rm CF}/\zeta$ 
via the spectral weight of 
Mott-Hubbard excitations $|4d_i^4, 4d_j^4\rangle \rightarrow  |4d_i^3, 4d_j^5\rangle$ 
between Ru sites $i$ and $j$.  
In the strong-coupling limit $t\ll U$ with hopping $t$ and intra-orbital Coulomb repulsion $U$, 
the dipole matrix element is proportional to $t$ and probes the same microscopic hopping processes that  
are relevant for magnetic exchange interactions. 
Based on selection rules, the optical spectral weight is sensitive to spin and orbital correlations between 
nearest neighbors \cite{Khaliullin04a,Oles05,Khaliullinrev,Fang03}, as observed in many 
$3d$ transition-metal compounds 
\cite{Miyasaka02,Kovaleva04,Lee05,Rauer06,GoesslingMn,GoesslingTi,Kovaleva10,Moskvin10,Reul12,Reul13}.
In a similar way, spin and orbital correlations determine the RIXS intensity of inter-site excitations 
studied at the O $K$ edge in $3d$ transition-metal oxides \cite{Monney13,Benckiser13}. 
\textit{A priori}, it is not clear in how far this strong-coupling approach based on local atomic multiplets 
works in $4d$ systems with larger bandwidth. 
However, optical data as well as electron energy loss spectroscopy (EELS) demonstrate the sensitivity 
to nearest-neighbor spin correlations in the $4d^5$ Kitaev material $\alpha$-RuCl$_3$ \cite{Sandilands16,Koitzsch20}. 
In Ca$_2$RuO$_4$ with $t_{2g}^4$ configuration, the optical conductivity $\sigma_1(\omega)$ shows 
a pronounced temperature dependence but due to different assignments of the observed features 
different conclusions were drawn for the orbital occupation \cite{Lee02,Jung03}. 
However, spin-orbit coupling has been neglected in the analysis thus far. 
We analyze the lowest Mott-Hubbard excitations using a local multiplet picture 
and show that the optical spectral weight can be used to estimate $\Delta_{\rm CF}/\zeta$. 
Furthermore, the optical excitation energies indicate a substantial temperature dependence of $\Delta_{\rm CF}$, 
in agreement with the temperature dependence of the tetragonal distortion of the 
RuO$_6$ octahedra \cite{Friedt01,Kunkemoeller17}, 
suggesting that $\Delta_{\rm CF}/\zeta$ is reduced at elevated temperature.

\section{Experimental methods}

Single crystals of Ca$_2$Ru$_{0.99}$Ti$_{0.01}$O$_4$ 
(space group $Pbca$) have been grown using the floating-zone method 
and were characterized by powder and single-crystal x-ray diffraction and measurements of 
the magnetization and the resistivity \cite{Kunkemoeller17}.  
At room temperature, the lattice constants are $a$\,=\,$5.4098(3)$\,\AA, $b$\,=\,$5.4683(4)$\,\AA, 
and $c$\,=\,$11.9781(9)$\,\AA{} \cite{Kunkemoeller17}, very similar to the values found in 
pristine Ca$_2$RuO$_4$ \cite{Braden98,Friedt01}. 
Ellipsometric data were measured on a polished $ab$ surface 
of a sample with $4\times 4\times 1$\,mm$^3$ in the energy range 0.75-5\,eV.\@ 
We use a rotating-analyzer ellipsometer (Woollam VASE) equipped with a retarder between polarizer and sample. 
Data were collected from 15\,K to 300\,K using a UHV cryostat with $p$\,$<$\,$10^{-9}$\,mbar. 
The angle of incidence equals 70$^\circ$. 
We corrected window effects by measuring a standard Si wafer for calibration. We studied an untwinned single crystal but could not resolve any anisotropy within the $ab$ plane. 
The $c$ axis is normal to the sample surface and contributes little to the data, 
the pseudo-dielectric function hence provides a reasonable estimate of the response 
within the $ab$ plane \cite{Aspnes80}. We demonstrate in Appendix A that considering 
the anisotropy explicitly yields a very similar result, in particular for the properties 
that we address in our theoretical calculations. In the analysis, a thin surface layer 
with a thickness of 4\,nm was included in order to account for the surface roughness.

\section{Optical conductivity}

\begin{figure}[t]
	\includegraphics[width=0.97\columnwidth,clip]{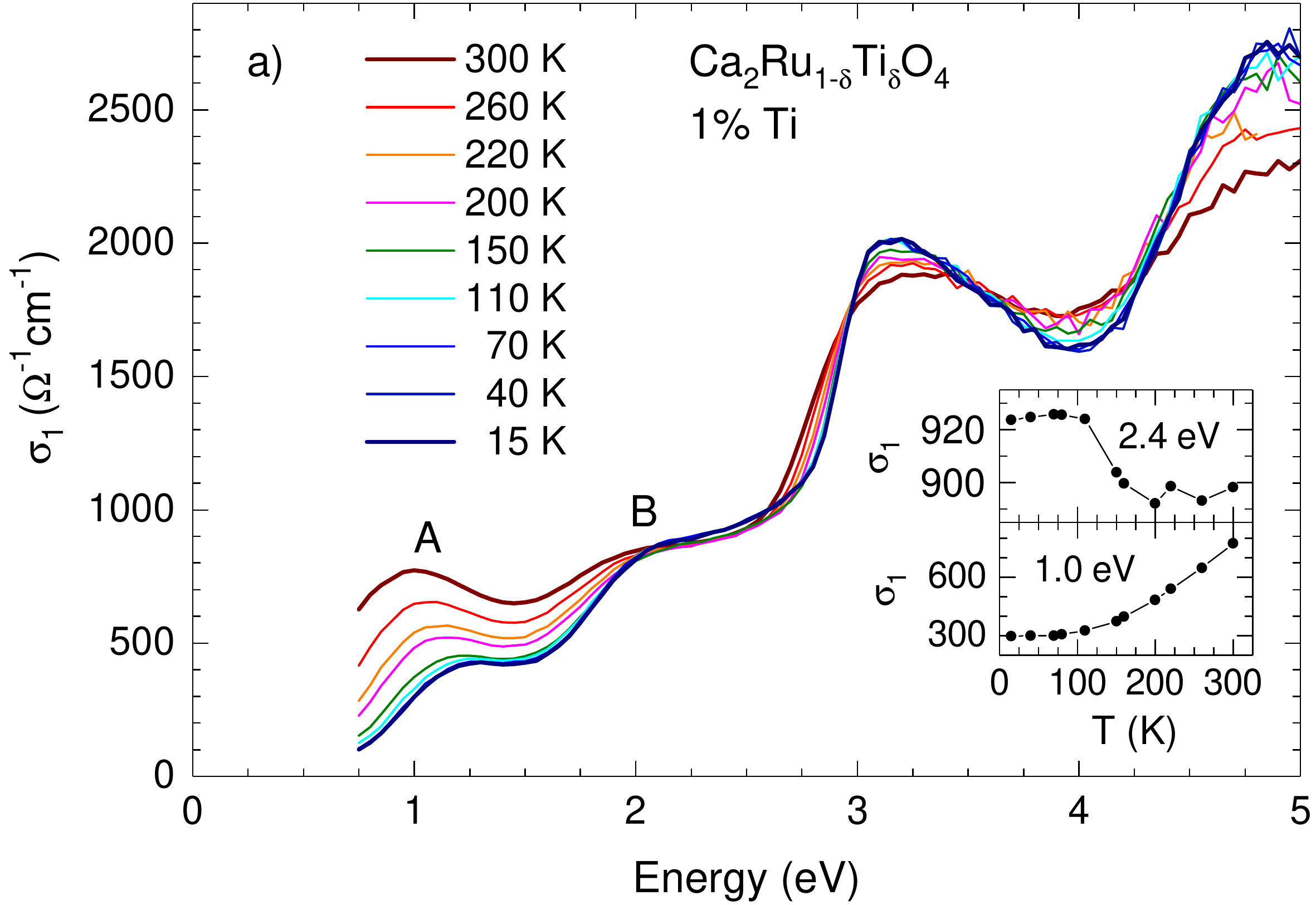}
	\includegraphics[width=0.97\columnwidth,clip]{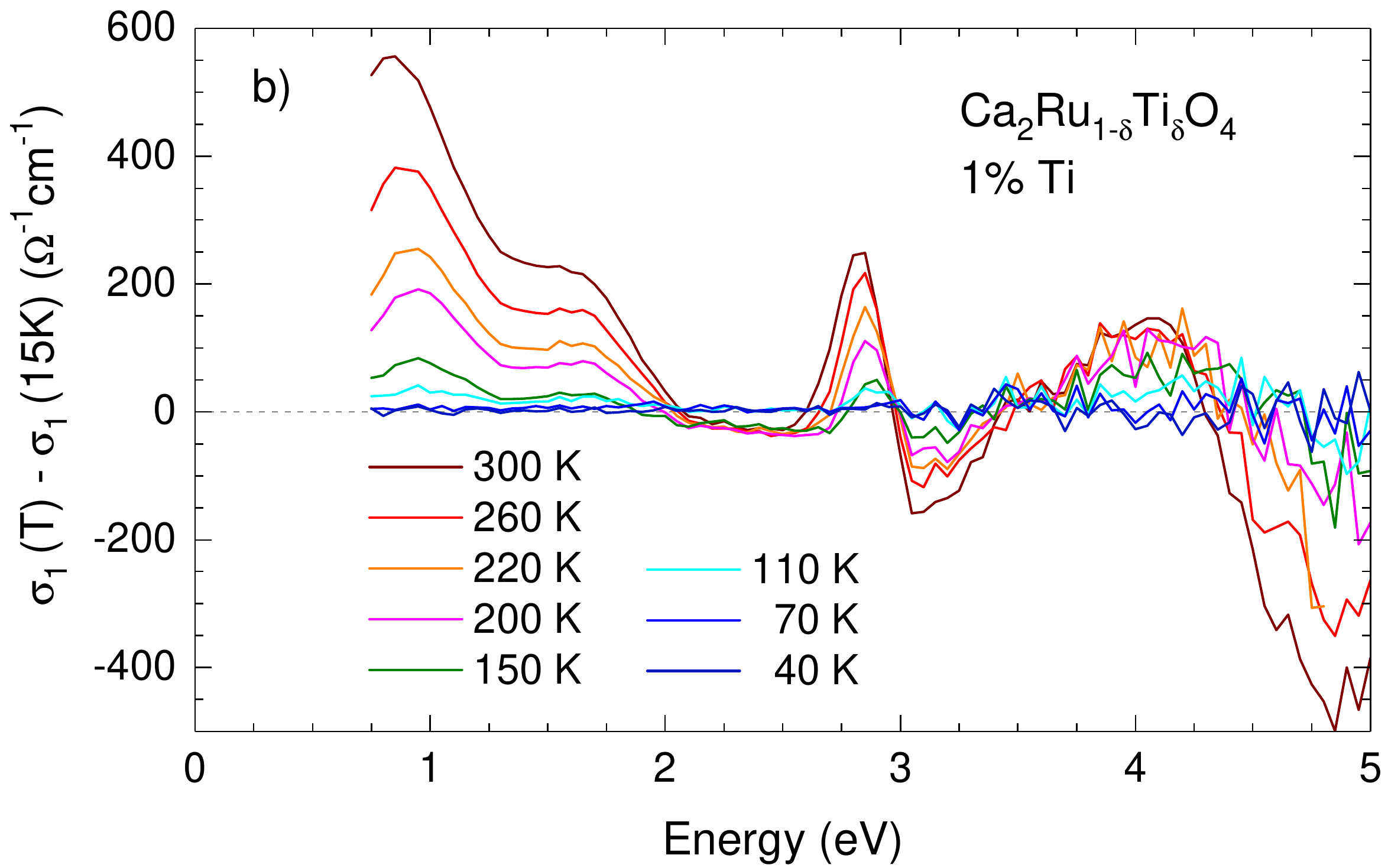}
	\caption{a) Optical conductivity of Ca$_2$Ru$_{0.99}$Ti$_{0.01}$O$_4$.  
	Insets: $\sigma_1$ at 1\,eV and 2.4\,eV.\@
	b) Difference spectra obtained by subtracting the 15\,K data. The spectral weight of peak A strongly increases with increasing temperature. Note that the shoulder at 1.2\,eV at high temperature coincides with the peak energy at low temperature, i.e., it reflects the energy shift of peak A rather than two separate features.  }
	\label{fig:sig1}
\end{figure}

The optical conductivity $\sigma_1(\omega)$ of Ca$_2$Ru$_{0.99}$Ti$_{0.01}$O$_4$ 
shows four peaks at about 1\,eV, 2\,eV, 3\,eV, and 5\,eV, see Fig.\ \ref{fig:sig1}a. 
Overall, our data agree very well with previous results for Ca$_2$RuO$_4$ 
based on a Kramers-Kronig analysis of reflectivity data \cite{Lee02,Jung03}. The substitution of 1\% of 
Ru$^{4+}$ ions by Ti$^{4+}$ ions has negligible effect on the optical properties, 
in agreement with the results obtained for other properties \cite{Kunkemoeller17}. 
Optical data of Ca$_2$RuO$_4$ were reported for 300\,K by Lee \textit{et al.}\,\cite{Lee02} and 
for 10\,K, 250\,K, 293\,K, 350\,K, and 370\,K by Jung \textit{et al.}\,\cite{Jung03}. 
The detailed temperature dependence, in particular in the vicinity of the N\'eel temperature 
$T_N$\,=\,110\,K \cite{Braden98}, has not been addressed thus far.

\begin{figure}[t]
	\includegraphics[width=0.9\columnwidth,clip]{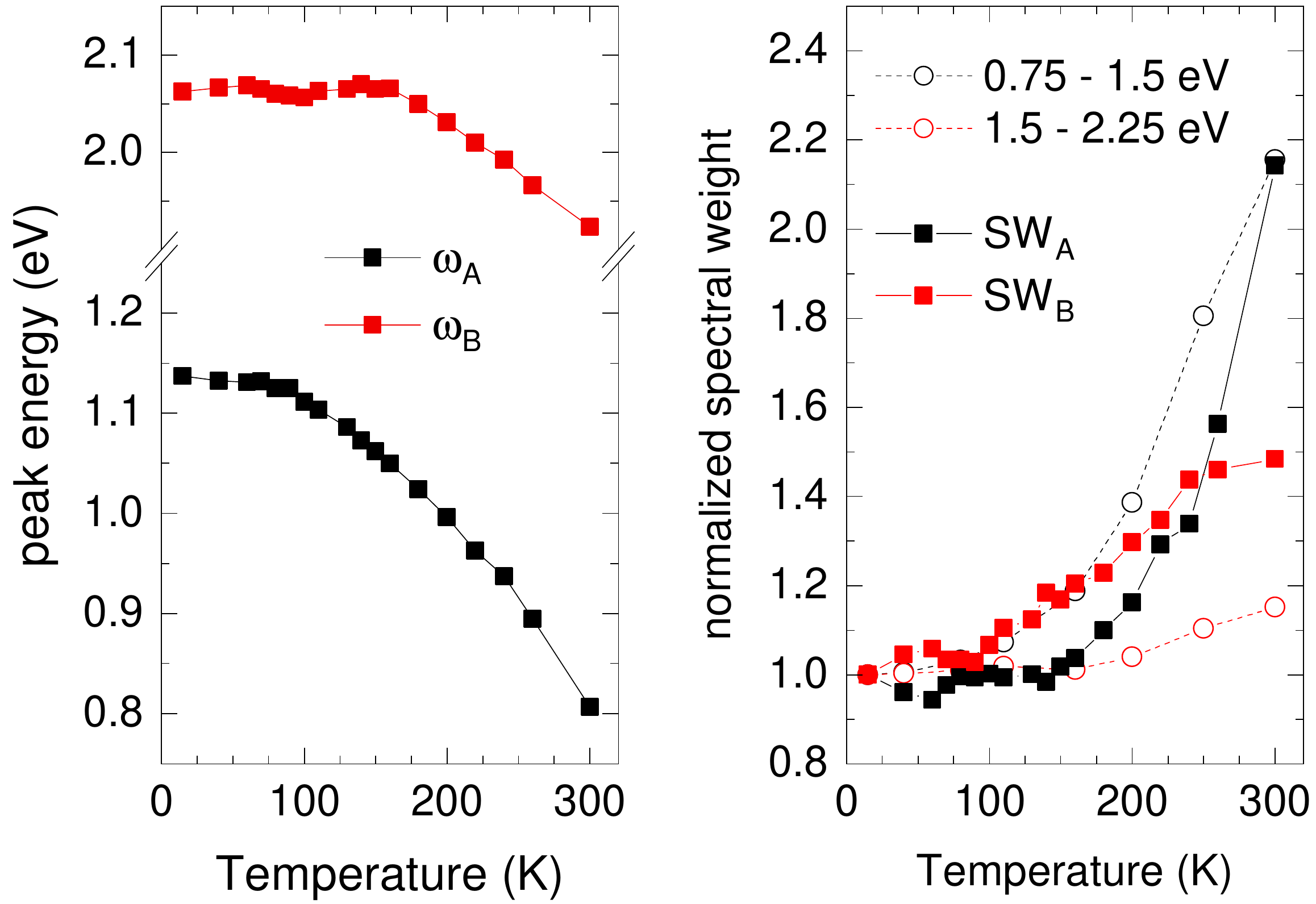}
	\caption{Left: Energies $\hbar \omega_i$ of peaks A and B as determined by an oscillator fit. 
		Right: Spectral weight of peaks A and B, normalized to the 15\,K value.  
		Full symbols show the fit result, open symbols correspond to the integral 
		of $\sigma_1(\omega)$ over the indicated ranges.
	}
	\label{fig:params}
\end{figure}

The analysis of the spectral weight is hampered if different excitations overlap. 
Therefore we focus on the two features with the lowest excitation energy, peaks A and B.\@ 
These are identified as Mott-Hubbard excitations, while charge-transfer excitations set in 
at about 3\,eV \cite{Lee02,Jung03}, see Sect.\ \ref{peakassignment}. 
To analyze the temperature dependence, we show $\sigma_1$ at 1.0\,eV and 2.4\,eV 
in the insets of Fig.\ \ref{fig:sig1}a and the difference spectra 
$\Delta \sigma_1$\,=\,$\sigma_1(T) - \sigma_1(15\,{\rm K})$ 
in Fig.\ \ref{fig:sig1}b. 
Peak A shows a remarkable increase of spectral weight with increasing temperature. 
Between 15\,K and 300\,K, its peak value increases by about a factor 2 while $\sigma_1(1\,{\rm eV})$ 
rises by almost a factor 3.
In contrast, $\sigma_1(2.4\,{\rm eV})$ shows a small step like decrease upon crossing $T_N$, 
characterizing the behavior in the range 2.0--2.5\,eV.\@ 
Furthermore, the data show a temperature-driven broadening of the charge-transfer gap around 3\,eV.\@

For a quantitative analysis, the complex ellipsometry data were fitted using an oscillator model. 
For peaks A and B, we employ two Gaussian oscillators. The asymmetric line shape of the charge-transfer gap 
around 3\,eV is well described by two Tauc-Lorentz oscillators, and two further Gaussians are used at higher energy. 
According to the fit, the spectral weight SW$_{\rm A}$ of peak A increases by a factor of about 2.2 between 
15\,K and 300\,K, see right panel of Fig.\ \ref{fig:params}, in agreement with previous results \cite{Jung03}. 
Moreover, the fit result is corroborated by direct integration of $\sigma_1(\omega)$ between 0.75\,eV 
and 1.5\,eV.\@ 
For peak B, the fit yields an increase of spectral weight $SW_{\rm B}$ by a factor 1.5, even though 
the peak value does not vary strongly with temperature. Also direct integration from 1.5\,eV to 2.25\,eV 
yields a weaker temperature dependence. The increase of the fit result for $SW_{\rm B}$ is mainly due to 
the enhanced width, making it more difficult to distinguish different features at high temperature. 

\begin{figure}[t]
	\includegraphics[width=\columnwidth,clip]{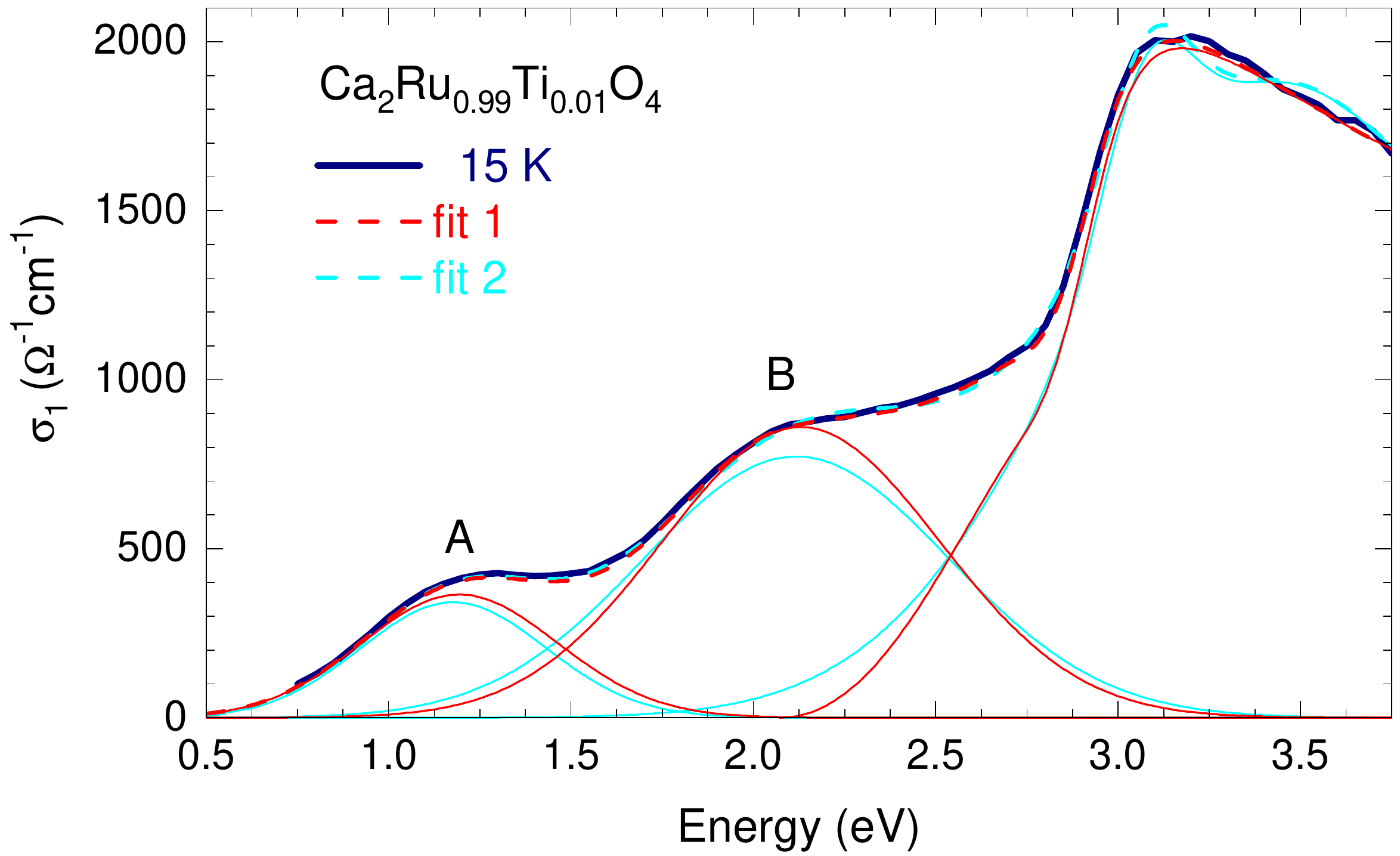}
	\caption{Two different oscillator fits of the 15\,K data. Fit 1 (red) uses two Tauc-Lorentz oscillators 
	for the 3\,eV feature, fit 2 (cyan) two Gaussian oscillators in the same range. Dashed lines show the 
	total fit, thin lines depict the contributions of the Gaussian oscillators for peaks A and B 
	as well as the sum of the other terms. For the spectral weight ratio SW$_{\rm A}$/SW$_{\rm B}$ 
	we find 0.30 in fit 1 and 0.27 in fit 2. 
}
	\label{fig:fit}
\end{figure}

For a comparison with theory (see below), the ratio SW$_{\rm A}$/SW$_{\rm B}$ is most interesting.  
We focus on the value at 15\,K, where the fit yields SW$_{\rm A}$/SW$_{\rm B}$\,$\approx$\,0.30. 
In detail, this result depends on the assumptions for the line shape of the strong feature at 3\,eV, 
see Fig.\ \ref{fig:fit}. 
Therefore, we performed a further fit in which the two Tauc-Lorentz oscillators were replaced by Gaussian 
oscillators. In this case, we find SW$_{\rm A}$/SW$_{\rm B}$\,$\approx$\,0.27. 
We hence consider SW$_{\rm A}$/SW$_{\rm B}$\,$\approx$\,1/3 to 1/4 at 15\,K.

The peak energies $\hbar\omega_i$ for $i$\,=\,A,B are depicted in the left panel of Fig.\ \ref{fig:params}. 
From 15\,K to 300\,K, $\omega_B$ changes by 7\,\% 
while peak A softens by about 0.3\,eV or 30\,\%. 
A similar change of 0.24\,eV was reported for peak A in Ca$_{2-\delta}$Sr$_{\delta}$RuO$_4$ for 
$\delta$\,=\,0.06 \cite{Lee02}. 
We will argue below that the strong temperature dependence of $\omega_A$ and 
$\Delta \omega$\,=\,$\omega_B-\omega_A$ reflect the pronounced reduction of the tetragonal 
crystal-field splitting $\Delta_{\rm CF}$.

\section{Peak assignment}
\label{peakassignment}

\subsection{Charge-transfer excitations}

We distinguish Mott-Hubbard excitations between Ru sites and charge-transfer excitations between 
Ru and O sites. The latter show larger spectral weight and, in Ca$_2$RuO$_4$, contribute above 
about 3\,eV \cite{Lee02,Jung03}, 
in agreement with angle-resolved photoelectron spectroscopy (ARPES) showing the onset of O bands 
about 2.5\,eV below the Fermi level \cite{Sutter17}. 
The charge-transfer peak at 3\,eV \cite{Lee02,Jung03} can be attributed to an electron transfer to the 
Ru $t_{2g}$ shell, $|$Ru\,$t_{2g}^4$, O\,$2p^6\rangle \rightarrow |$Ru\,$t_{2g}^5$, O\,$2p^5\rangle$. 
The peak at 5\,eV predominantly corresponds to excitations into the Ru $e_g$ shell
to a $|$Ru\,$t_{2g}^4 e_g^1$, O\,$2p^5\rangle$ final state.
Compared to the $t_{2g}^5$ peak at 3\,eV, the larger spectral weight agrees with the 
enhanced hopping matrix elements between O $2p$ and Ru $e_g$ states. 
The energy difference of about 2\,eV between the peaks in $\sigma_1(\omega)$ 
is similar to the $t_{2g} - e_g$ splitting observed in x-ray absorption or 
resonant inelastic x-ray scattering \cite{Fatuzzo15,Das18,Gretarsson19}.

\subsection{Mott-Hubbard excitations}

For the assignment of the Mott-Hubbard excitations  
$|d_i^4, d_j^4\rangle \rightarrow |d_i^3, d_j^5\rangle$ between Ru sites $i$ and $j$, 
we address the $d^3$, $d^4$, and $d^5$ states in a local multiplet scenario \cite{Zhang17,Oles05}. 
The cubic crystal-field splitting $10$\,Dq is large enough to break Hund's rule. 
We thus may neglect the $e_g$ orbitals, all four electrons occupy the $t_{2g}$ shell in 
the $d^4$ ground state as well as in the lowest excited states with $d^3$ and $d^5$ multiplets. 
For the assignment of peaks A and B at 1\,eV and 2\,eV, it is sufficient to consider the Coulomb interaction 
within the $t_{2g}$ shell, which can be described by two parameters, the intra-orbital Coulomb repulsion $U$ 
and Hund coupling $J_H$ \cite{Tanabe,Georges13}. 
In cubic approximation, the $d^4$ ground state is given by the $S$\,=\,1 multiplet $^3T_1$ at 
$6U-13 J_H$, see Tab.\ \ref{tab:multipletenergies}. 
At this stage, we may neglect tetragonal distortion and spin-orbit coupling, which will be 
considered below for the quantitative analysis.

The Mott-Hubbard excitation energies of the different $|d_i^3, d_j^5\rangle$ excited states are given by 
\begin{eqnarray}
\label{eq:energyMH} \nonumber
E_{{\rm MH},m} \! & = & \! E(t_{2g}^3(m))+E(t_{2g}^5(^2T_2))-2E(t_{2g}^4(^3T_1)) 
\\ 
& = & \! E(t_{2g}^3(m)) -2U + 6 J_H     \,\,  .
\end{eqnarray}
Since all six $t_{2g}^5$ configurations belong to the same cubic $^2T_2$ multiplet, the excitation energies
$E_{{\rm MH},m}$ are determined by the four $t_{2g}^3$ multiplets $^4A_2$, $^2T_1$, $^2E$, and $^2T_2$. 
Since $^2T_1$ and $^2E$ are degenerate, there are only three different values of $E_{{\rm MH},m}$, 
see Tab.\ \ref{tab:multipletenergies}.

In cubic approximation, peaks A and B hence correspond to excitations to 
$|d_i^3(^4\!A_2), d_j^5(^2T_2)\rangle$ and $|d_i^3(^2T_1/^2E), d_j^5(^2T_2)\rangle$, respectively. 
This assignment is supported by LDA+DMFT calculations \cite{Zhang17} which obtain good agreement 
with other experimental results for two different parameter sets, either $U$\,=\,2.3\,eV and 
$J_H$\,=\,0.4\,eV or $U$\,=\,3.1\,eV and $J_H$\,=\,0.7\,eV.\@ 
The former set yields 1.1\,eV and 2.3\,eV for peaks A and B in $\sigma_1(\omega)$, in good agreement 
with our data. We neglect the $t_{2g}^3(^2T_2)$ multiplet since the corresponding excitation energy 
$U+2J_H$ lies above 3\,eV, the absorption band hence overlaps with charge-transfer excitations. 
In Ca$_2$RuO$_4$, the applicability of the local multiplet picture is further supported by ARPES \cite{Sutter17}, 
where the electron removal states closest to the Fermi energy $E_F$ have been found to show $t_{2g}^3$ character. 
At 150\,K, ARPES finds two flat bands at 0.8\,eV and 1.7\,eV below $E_F$ \cite{Sutter17}. 
The former has been attributed to the $^4\!A_2$ multiplet, and the energy difference of 0.9\,eV between 
the $t_{2g}^3$ ARPES bands agrees with the splitting between peaks A and B in our optical data, in agreement 
with Eq.\ \ref{eq:energyMH}. 
The flatness of the ARPES bands explains the observation of well-defined peaks in $\sigma_1(\omega)$. 
For the band at 1.7\,eV below $E_F$, the flat character was attributed to 
the strong anisotropy in spin space and the quasi-1D character of $xz/yz$ bands \cite{Klosinski20}. 

A third ARPES band lying about 2\,eV below $E_F$ has been attributed to the $^2E$ multiplet \cite{Sutter17}. 
It exhibits a larger dispersion, suggesting the existence of a broad feature at about 2.5\,eV in $\sigma_1(\omega)$. 
However, our theoretical analysis below shows that the $^2E$ multiplet carries much less spectral weight 
in $\sigma_1(\omega)$ than the $^2T_1$ multiplet. The weakness and large width explain the absence of 
a clear feature in our data.

Our peak assignment further agrees with the results of Jung \textit{et al}.\ \cite{Jung03}, 
who combined optical spectroscopy with an LDA+$U$ study neglecting spin-orbit coupling (see also \cite{Fang04}). 
A different assignment was favored by Lee \textit{et al}.\ \cite{Lee02} but they neglect rotation and tilt 
of the octahedra. In this case, the matrix elements for peak A vanish. 
We will address the spectral weight and in particular the effect of spin-orbit coupling 
in Sect.\ \ref{sect:SW}.

\begin{table}[t]
	\begin{tabular}{ccccc}
		multiplet & \hspace{2mm} & energy \cite{Zhang17} & \hspace{2mm} & $E_{{\rm MH},m}$ \\ 
		\hline 
		$t_{2g}^4(^3T_1)$	    & &  $6U-13J_H$ & & \\
		$t_{2g}^5(^2T_2)$	    & & $10U-20J_H$ & & \\
		$t_{2g}^3(^4\!A_2)$	    & &  $3U- 9J_H$ & & $U-3J_H$  \\
		$t_{2g}^3(^2T_1/^2E)$	& &  $3U- 6J_H$ & & $U$   \\
		$t_{2g}^3(^2T_2)$	    & &  $3U- 4J_H$ & & $U+2J_H$  
	\end{tabular} 
	\caption{Cubic multiplets relevant for peaks A and B and their energies in terms of the intra-orbital 
		Coulomb repulsion $U$ and Hund coupling $J_H$ \cite{Zhang17,Georges13,CommentJH}. 
		The third column gives the corresponding Mott-Hubbard excitation energy, see Eq.\ (\ref{eq:energyMH}).   
	}
	\label{tab:multipletenergies}
\end{table}

\section{Tetragonal crystal field}

The energy difference $\hbar\Delta \omega$ between peaks A and B increases from 0.9\,eV at 15\,K to 1.1\,eV
at 300\,K, see Fig.\ \ref{fig:params}. 
In cubic approximation, one expects a temperature-independent value, $\hbar\Delta \omega$\,=\,$3J_H$. 
The experimental result can be rationalized by taking into account the temperature-dependent distortion of the RuO$_6$ octahedra \cite{Friedt01}. 
The tetragonal crystal field lowers the $xy$ orbital by $\Delta_{\rm CF}$ 
with respect to $xz$ and $yz$ and lifts the degeneracy of the $t_{2g}^4(^3T_1)$ multiplet. 
This yields $xy$ orbital order, the local $d^4$ ground state being the spin triplet $|xy^2,S_z\rangle$ 
(spin $S$\,=\,1, $S_z$\,=\,$\pm 1,0$) with double occupancy of the $xy$ orbital \cite{Zhang17}, 
\begin{eqnarray}
\label{eq:xy2}
|xy^2,\,1\rangle & = & c^\dagger_{xz\uparrow} c^\dagger_{yz\uparrow} c^\dagger_{xy\uparrow} 
c^\dagger_{xy\downarrow} |0\rangle
\\ \nonumber
|xy^2,\,0\rangle & = & \frac{1}{\sqrt{2}}\left(   
c^\dagger_{xz\uparrow} c^\dagger_{yz\downarrow} + c^\dagger_{xz\downarrow} c^\dagger_{yz\uparrow}
\right) c^\dagger_{xy\uparrow} c^\dagger_{xy\downarrow} |0\rangle
\\ \nonumber
|xy^2,\,-1 \rangle & = & c^\dagger_{xz\downarrow} c^\dagger_{yz\downarrow} c^\dagger_{xy\uparrow} c^\dagger_{xy\downarrow} |0\rangle \, ,
\end{eqnarray}
where $|0\rangle$ denotes the vacuum state and $c^\dagger_{\tau\sigma}$ creates an electron in orbital $\tau$ 
with spin $\sigma$.

\begin{figure}[tb]
	\includegraphics[width=0.85\columnwidth,clip]{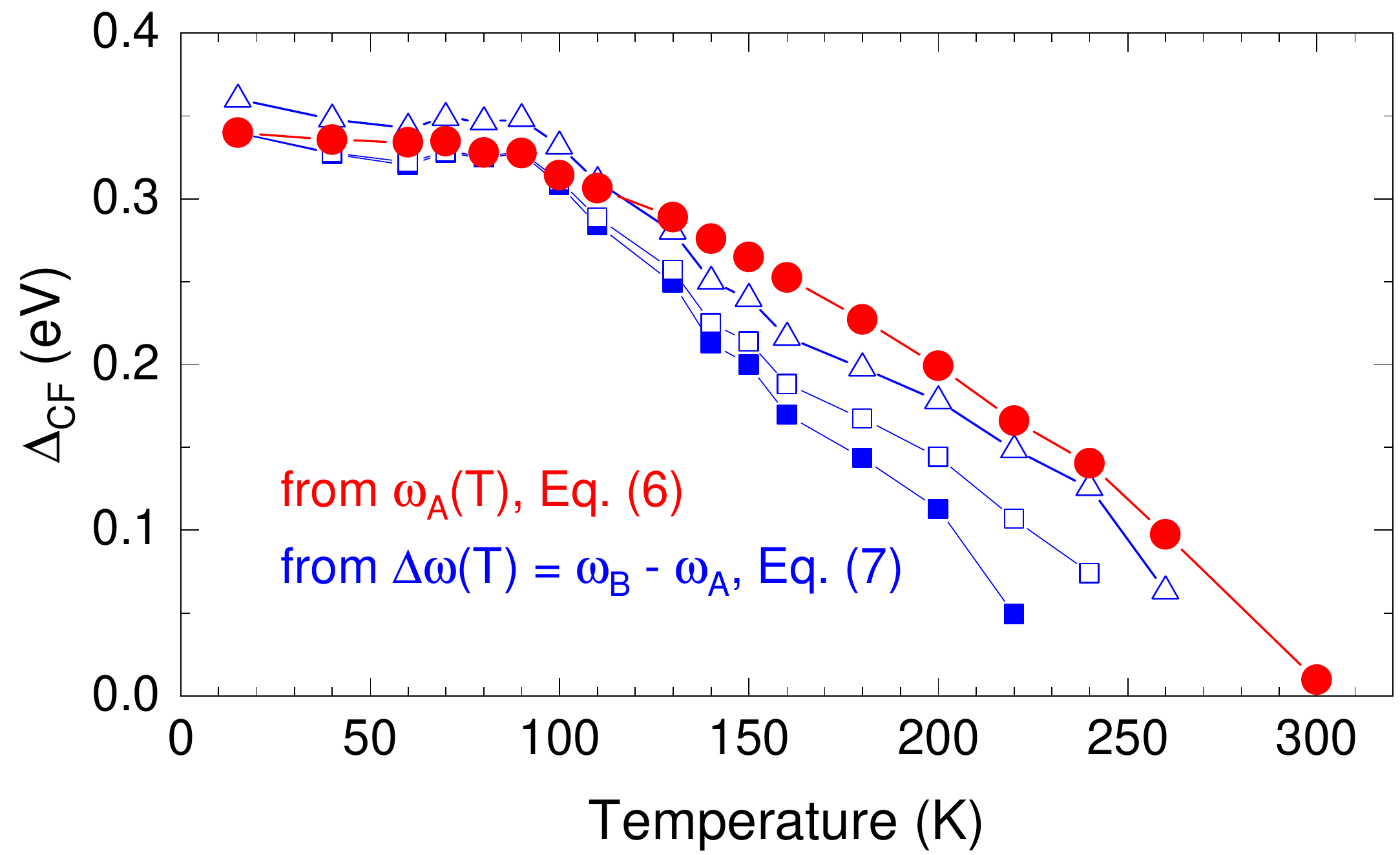}
	\caption{Tetragonal crystal-field splitting $\Delta_{\rm CF}(T)$ based on Eq.\,(\ref{eq:Delta2}) (red) 
	for $\Delta_{\rm CF}(15\,K)$\,=\,0.34\.eV \cite{Das18,Gretarsson19}. 
	Blue symbols denote the result of Eq.\ (\ref{eq:Delta}) for $J_H$\,=\,0.4\,eV (full) or 0.35\,eV (open) 
	with $\Delta_{\rm CF}(15\,K)$\,=\,0.34\.eV (squares) or 0.36\,eV (triangles). 
	}
	\label{fig:Delta}
\end{figure}

The tetragonal field also splits the $t_{2g}^5(^2T_2)$ multiplet but this does not yield a splitting in 
$\sigma_1(\omega)$ since adding an electron to $|xy^2,\,S_z\rangle$ only yields those $t_{2g}^5$ states 
with doubly occupied $xy$ orbital. All of them have the same energy. 
Considering the $d^3$ states, the $^4\!A_2$ and $^2\!E$ multiplets are not split by a tetragonal field 
but $^2 T_1$ is. 
The three lowest Mott-Hubbard excitation energies then are 
\begin{eqnarray}
\label{eq:EA}
E_{\rm A} & = & U - 3J_H + \Delta_{\rm CF} \\
E_{\rm B} & = & U + \Delta_{\rm CF} + J_H - \sqrt{\Delta_{\rm CF}^2 + J_H^2} \\
E_{\rm C} & = & U + \Delta_{\rm CF}  \, .
\end{eqnarray}
Since $U$ and $J_H$ can be viewed as being independent of temperature, this yields several ways to estimate 
$\Delta_{\rm CF}(T)$. 
From Eq.\ (\ref{eq:EA}) we obtain 
\begin{equation}
\label{eq:Delta2}
\Delta_{\rm CF}(T) = \Delta_{\rm CF}(15\,{\rm K}) + \hbar\omega_{\rm A}(T) - \hbar\omega_{\rm A}(15\,{\rm K}) \, , 
\end{equation}
which is plotted in Fig.\ \ref{fig:Delta} for $\Delta_{\rm CF}(15\,K) \approx 0.34$\,eV as determined from 
the peak energy observed in RIXS \cite{Das18,Gretarsson19}. 
The RIXS peak directly yields $\Delta_{\rm CF}$ if we assume $\zeta$\,=\,0. For finite $\zeta$, the excitation 
energy is slightly larger than $\Delta_{\rm CF}$, see Fig.~\ref{fig:energiesLS}. 
Below we establish $\Delta_{\rm CF}/\zeta \geq 2.4$ as a lower bound, in which case the average excitation energy amounts to $\lesssim 1.1$\,$\Delta_{\rm CF}$.

The RIXS data show little change between 16\,K and 125\,K \cite{Das18}, in agreement with our result. 
Additionally, we find that $\Delta_{\rm CF}$ is strongly suppressed at 300\,K, in agreement with the 
strongly reduced distortion of the RuO$_6$ octahedra that changes from compression to elongation at about 
300\,K \cite{Friedt01,Kunkemoeller17}. 
The strong suppression of $\Delta_{\rm CF}$ also agrees with the pronounced change of the orbital occupation observed in x-ray absorption \cite{Mizokawa01,Pincini19}. 
Furthermore, resonant elastic scattering suggests that the orbital polarization 
vanishes close to 300\,K \cite{Zegkinoglu05}. 
Note, however, that undistorted octahedra do not imply 
that $\Delta_{\rm CF}(300\,{\rm K})$ vanishes exactly because we also have to consider the crystal-field 
contributions of further neighbors in the layered crystal structure.
  
For a second estimate, we identify the experimental peak splitting 
$\Delta\omega$\,=\,$\omega_{\rm B}-\omega_{\rm A}$ with $E_B-E_A$ and find
\begin{eqnarray}
\label{eq:Delta}
\nonumber
\Delta^2_{\rm CF}(T) & = & \bigg[ \sqrt{\Delta_{\rm CF}^2(15\,{\rm K})+J_H^2}+\hbar\Delta\omega(15\,{\rm K}) 
	\\
	& &   - \hbar\Delta\omega(T) \bigg]^2-J_H^2  \, . 
\end{eqnarray}
The result agrees well with that of Eq.\ (\ref{eq:Delta2}) for $J_H$\,=\,0.35\,eV, see open symbols 
in Fig.\ \ref{fig:Delta}. 
First-principles calculations \cite{Zhang17} 
find $\Delta_{\rm CF}$\,$\approx$\,0.3\,eV in the insulating phase at 180\,K and 0.1\,eV in the metallic phase at 400\,K.\@ 
Considering spin-orbit coupling and further on-site $4d^4$ excitation energies, 
$L$ edge RIXS results yield $\Delta_{\rm CF}$\,=\,0.25\,eV and 
$\zeta$\,=\,0.13\,eV at 12\,K \cite{Gretarsson19}.

\section{Spectral weight and spin-orbit coupling}
\label{sect:SW}

Spin-orbit coupling $\zeta$\,=\,$2\lambda$ is of the order of 0.1\,eV in Ca$_2$RuO$_4$ \cite{Zhang17,Gretarsson19} 
and thus can be neglected for the assignment of broad peaks observed at 1\,eV and 2\,eV.\@ 
However, it plays a decisive role for the spectral weight of these peaks since it affects the orbital occupation, 
lifting the degeneracy of the $|xy^2,S_z\rangle$ ground state. 
In the following, we first calculate the spectral weight for a finite tetragonal crystal-field splitting 
$\Delta_{\rm CF}$ with $\zeta$\,=\,0. In a second step, we study finite $\zeta$.

\subsubsection{Double occupancy of $xy$ for $\zeta$\,=\,0}

The optical matrix elements for Mott-Hubbard excitations depend on the hopping amplitudes 
between nearest-neighbor Ru sites and hence reflect spin and orbital correlations, see Appendix B.\@  
The restriction to nearest-neighbor sites is justified since hopping to further neighbors is small. 
In our optical data, we could not resolve any anisotropy within the $ab$ plane. 
We therefore treat all nearest-neighbor Ru-Ru pairs as equivalent and consider a Ru-O-Ru bond 
along the global $x$ axis. On each of the two Ru sites $k \in \{i,j\}$, we employ a local reference frame 
$(x_k,y_k,z_k)$ where the local axes point from the central Ru ion towards the O ligands. 
The effective Ru-Ru hopping matrix from $(xy_i,yz_i,xz_i)$ to $(xy_j,yz_j,xz_j)$ reads
\begin{equation}
\label{eq:tmat}
\hat{t} = \frac{t_{pd\pi}^2}{\Delta_{\rm CT}}\left(   
\begin{array}{ccc}
\alpha_{xy}  & 0  & -\beta \\ 
0 & 0  &  0 \\ 
\beta & 0  & \alpha_{xz} 
\end{array} 
\right) \, ,
\end{equation}
where $t_{pd\pi}$ denotes Ru-O hopping and $\Delta_{\rm CT}$ is the charge transfer energy. 
Using the local reference frames has the advantage to yield zero hopping for all processes 
that involve $yz_k$ orbitals. 
The coefficients $\alpha_{u}$ and $\beta$ depend on the octahedral tilt and rotation angles 
$\theta$ and $\varphi$ which are taken into account by a rotation on the O ion from one reference frame 
to the other. 
For a 180$^\circ$ bond one obtains $\alpha_{xy}$\,=\,$\alpha_{xz}$\,=\,1 and $\beta$\,=\,0.
In Ca$_2$RuO$_4$, both $\theta$ and $\varphi$ roughly equal 11$^\circ$ and show only a small change 
as a function of temperature \cite{Braden98}. 	
This yields $\alpha_{xy}$\,$\approx$\,0.88, $\alpha_{xz}$\,$\approx$\,0.95, and 
$\beta$\,$\approx$\,0.31.
For simplicity, we employ $\alpha_{xy}$\,=\,$\alpha_{xz}$\,=\,1 and $\beta$\,=\,1/3.

For double occupancy of the $xy$ orbital and AF order, the hopping matrix yields the optical matrix elements  
\begin{eqnarray}
\label{eq:4a2AF}
|M(^4\!A_2)|^2   & = & \frac{2}{3}\beta^2 \, \left(t_{pd\pi}^2/\Delta_{\rm CT} \right)^2   \\
\label{eq:2T1AF}
|M(^2\!T_1)|^2   & = & \left(t_{pd\pi}^2/\Delta_{\rm CT}\right)^2    \\
\label{eq:2EAF}
|M(^2\!E)|^2     & = & \frac{4}{3}\beta^2 \, \left(t_{pd\pi}^2/\Delta_{\rm CT}\right)^2   
\end{eqnarray}
where we neglect the small spin canting of about 3.5$^\circ$ \cite{Kunkemoeller17} and treat spins 
on neighboring sites as being antiparallel.
The spectral weight of the $^2\!E$ multiplet vanishes in the absence of octahedral tilt and rotation, 
i.e., $\beta$\,=\,0. Therefore, its contribution to $\sigma_1(\omega)$ is about an order of magnitude 
smaller than the one of the $^2T_1$ multiplet. This may explain why we cannot resolve a weak separate 
contribution of the $^2E$ states, in contrast to ARPES \cite{Sutter17}. 

The spectral weight of the $^4\!A_2$ multiplet, which corresponds to peak A in $\sigma_1(\omega)$, 
also vanishes for $\beta$\,=\,0. For comparison with experiment, we are mainly interested in the 
spectral weight ratio $SW_{\rm A}/SW_{\rm B}$. 
For $\beta$\,=\,1/3 and AF order at $T$\,=\,0, 
Eqs.\ (\ref{eq:4a2AF}) and (\ref{eq:2T1AF}) predict 
$SW_{\rm A}/SW_{\rm B}$\,=\,$\frac{2}{3}\beta^2 \, \omega_{\rm B}/\omega_{\rm A}$\,$\approx $\,0.13, 
in clear disagreement with the experimental result $1/3$ to $1/4$. 
This discrepancy can be resolved by considering finite spin-orbit coupling $\zeta$, as discussed below.

The spectral weight reflects nearest-neighbor spin and orbital correlations. 
In the limit $\zeta$\,=\,0 with $\Delta_{\rm CF} > 0$, the local ground state is given by 
Eq.\ (\ref{eq:xy2}) and the orbital occupation is independent of the size of $\Delta_{\rm CF}$. 
In this limit, the temperature dependence of the spectral weight reflects changes of the nearest-neighbor 
spin-spin correlations. In layered Ca$_2$Ru$_{0.99}$Ti$_{0.01}$O$_4$, 
we do not expect pronounced changes of the spectral weight 
at $T_N$. In two-dimensional materials, the spin correlation length remains large above the 
three-dimensional ordering temperature, and both the correlation length and the nearest-neighbor 
spin-spin correlations drop slowly with increasing temperature. The corresponding slow change of 
the spectral weight was observed in the layered $3d^4$ compound LaSrMnO$_4$ \cite{GoesslingMn}. 
Without a detailed theoretical prediction for the behavior of the spin-spin correlations, 
we have to compare the results for the AF ordered state at $T$\,=\,0 given in 
Eqs.\ (\ref{eq:4a2AF}) and (\ref{eq:2T1AF}) with those for a magnetically fully disordered state 
at $T\gg T_N$, neglecting any change of the orbital occupation. 
We emphasize that the octahedral tilt and rotation angles and hence $\beta$ do not vary strongly 
with temperature \cite{Braden98}. 
For the $^4\!A_2$ multiplet we find an increase of spectral weight by a factor 2, similar to our 
experimental result for peak A.\@ 
However, $^2T_1$ and $^2\!E$ exhibit the opposite behavior, a suppression by a factor 2, 
in contrast to the optical data, see inset of Fig.\ \ref{fig:sig1}a and Fig.\ \ref{fig:params}. 
Our predictions roughly agree with LDA+$U$ results for peaks A and B which also neglect 
spin-orbit coupling \cite{Jung03,Fang04}. 
Such opposite behavior of the lowest two absorption bands indeed was observed in the $3d^2$ compound
YVO$_3$ \cite{Reul12}, an electron analogue to the $d^4$ configuration but with the three-dimensional 
perovskite structure. Note that spin-orbit coupling is negligible for the $3d$ vanadates.  
We conclude that neither the experimental value of $SW_{\rm A}/SW_{\rm B}$ at 15\,K nor the 
temperature dependence of $SW_{\rm B}$ can be described in a scenario that neglects spin-orbit coupling.

\subsubsection{Finite spin-orbit coupling}

Spin-orbit coupling affects the orbital occupation and hence the spectral weights. 
In the following, we first discuss the effect of finite $\zeta$ on the electronic states 
and then address the spectral weight as a function of $\Delta_{\rm CF}/\zeta$.

In the limit $\Delta_{\rm CF}$\,=\,0, i.e., cubic symmetry, the $^3T_1$ ground state shows 
spin $S$\,=\,1 and $L_{\rm eff}$\,=\,1. These are coupled to $J$\,=\,0, 1, and 2, and the corresponding 
states have the energies $-\zeta$, $-\zeta/2$, and $+\zeta/2$, respectively, see Fig.\ \ref{fig:energiesLS}. 
In the opposite limit, $\zeta$\,=\,0 and positive $\Delta_{\rm CF}$, the ground state is a spin triplet 
with double occupancy of the $xy$ orbital, cf.\ Eq.\ (\ref{eq:xy2}). 
In between these two limits, the local ground state for a single $d^4$ site is given by the singlet $|s\rangle$ \cite{Sarvestani18,Jain17},  
\begin{eqnarray}
\label{eq:sing}
|s\rangle   & = &  \cos\theta_s \, |xy^2, 0\rangle \\ 
\nonumber
& + & \frac{\sin\theta_s}{2} \left(-|yz^2,1\rangle \! + \! |yz^2,-1\rangle 
\! + \!  i |xz^2,1\rangle \! + \! i |xz^2,-1\rangle\right) 
\end{eqnarray}
and the lowest excited state is the doublet $|d_{\pm 1}\rangle$, 
\begin{align}
\label{eq:doup1}
|d_{+1}\rangle  &  =  \cos\theta_d \, |xy^2, 1\rangle 
+ \frac{\sin\theta_d}{\sqrt{2}} ( i |xz^2,0\rangle + |yz^2,0\rangle)
\\
\label{eq:doum1}
|d_{-1}\rangle  &  =  \cos\theta_d \, |xy^2,-1\rangle 
\! + \! \frac{\sin\theta_d}{\sqrt{2}} ( i  |xz^2,0\rangle \!- \! |yz^2,0\rangle )
\end{align}
where the prefactors depend only on 
$\Delta_{\rm CF}/\zeta$ \cite{Jain17}, 
\begin{eqnarray}
\nonumber
\tan\theta_d & = & \left( (\Delta_{\rm CF}/\zeta) + 
\sqrt{1+(\Delta_{\rm CF}/\zeta)^2}\right)^{-1} \\
\nonumber
\tan\theta_s & = & \sqrt{1+\gamma^2}-\gamma \,\,\,\,\,\, \textrm{for} \,\,\, \gamma = \left((\Delta_{\rm CF}/\zeta)-1/2\right)/\sqrt{2} \, . 
\end{eqnarray}
For $\zeta$\,=\,$0$ one finds $\cos\theta_s$\,=\,$\cos\theta_d$\,=\,1, which yields the states 
$|xy^2, S_z\rangle$ described in Eq.\ (\ref{eq:xy2}). 
The other states $|yz^2, S_z\rangle$ and $|xz^2, S_z\rangle$ correspond to double occupancy of the $yz$ or 
$xz$ orbital, respectively, and are constructed equivalently, cf.\ Eq.\ (\ref{eq:xy2}) \cite{Zhang17}. 
The ratio $\Delta_{\rm CF}/\zeta$ determines the orbital occupation and thereby strongly affects 
the spectral weight.

\begin{figure}[t]
	\includegraphics[width=0.9\columnwidth,clip]{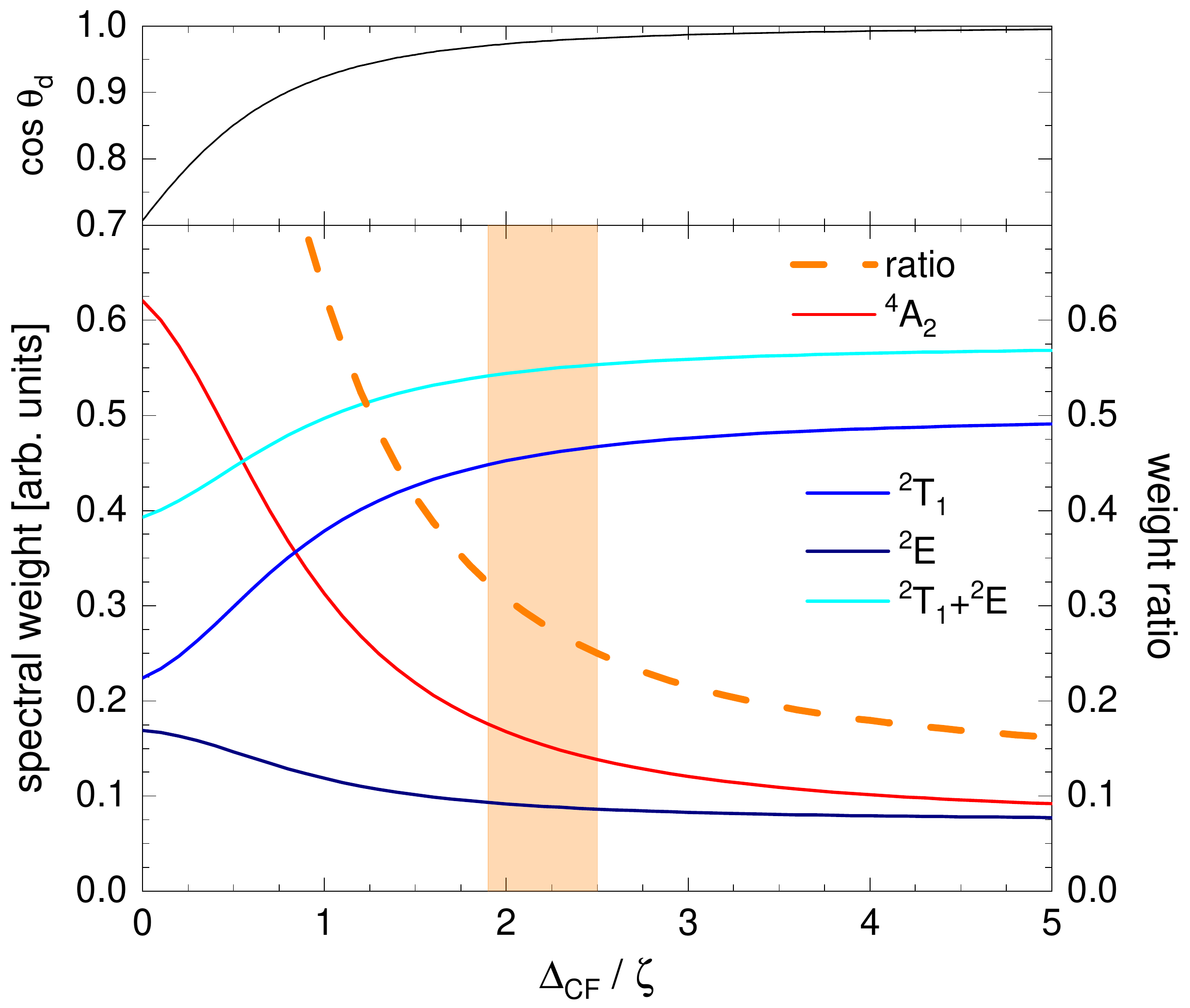}
	\caption{Results for the doublet ground state $|d_{+1}\rangle_i |d_{-1}\rangle_j$ 
	which neglects any singlet contribution.
	Top: Weight factor $\cos\theta_d$ for double occupancy of the $xy$ orbital,  
	cf.\ Eqs.\ (\ref{eq:doup1}) and (\ref{eq:doum1}).
	Bottom, left axis: Spectral weight of peak A ($^4A_2$ multiplet, red) and of the 
	two contributions to peak B ($^2T_1$ and $^2E$, blue) as a function of 
	$\Delta_{\rm CF}/\zeta$. 
	Right axis (dashed orange line): spectral weight ratio ${\rm SWR}$\,=\,$SW({}^4\!A_2)/SW(^2T_1+{} ^2\!E)$. 
	The shaded area denotes the range $1/4 \leq {\rm SWR} \leq 1/3$.
	}
	\label{fig:SWratio}
\end{figure}

The energies of singlet and doublet are given by 
\begin{align}
E_s/(\zeta/2) & = -\frac{\Delta_{\rm CF}}{3\,\zeta} -\frac{1}{2} 
-\sqrt{\left(\frac{\Delta_{\rm CF}}{\zeta} -\frac{1}{2}\right)^2+2} 
\end{align}
\begin{align}
E_d/(\zeta/2) & =  -\frac{\Delta_{\rm CF}}{3\,\zeta}
-\sqrt{\left(\frac{\Delta_{\rm CF}}{\zeta}\right)^2 +1} 
\end{align}
and their splitting amounts to 
$E_{sd} \approx (\zeta/2)^2/\Delta_{\rm CF}$ in the limit $\zeta \ll \Delta_{\rm CF}$. 
The local non-magnetic singlet ground state obtained for any finite value of $\zeta$ appears to be at odds 
with the occurrence of magnetic order below $T_N$. 
Starting from this single-site singlet picture, however, van-Vleck-type magnetism can be favored on the lattice by exchange 
interactions between excited states \cite{Khaliullin13}. 
Antiferromagnetic order with magnetic moments pointing within the $ab$ plane, as observed experimentally, is achieved via condensation of \cite{Akbari14,Feldmaier20,Strobel21} 
\begin{equation}
\label{eq:dxy}
|d_{x/y}\rangle  = \frac{1}{\sqrt{2}}	\left( |d_{+1}\rangle \pm |d_{-1}\rangle    \right) \, ,
\end{equation}
where the + (-) sign refers to $|d_x\rangle$ ($|d_y\rangle$).
The actual local ground state is a mixture of $|d_{x/y}\rangle$ and singlet $|s\rangle$ 
which in particular depends on hopping interactions \cite{Feldmaier20,Strobel21}. 
A determination of the precise ground state on the lattice is beyond the scope of our study but it turns out that this is not necessary to obtain a reliable estimate of $\Delta_{\rm CF}/\zeta$.
First of all we find the lower bound $\Delta_{\rm CF}/\zeta > 2.2 \pm 0.3$ if we neglect any singlet contribution and consider $|d_{+1}\rangle_i |d_{-1}\rangle_j$ as ground state for the calculation 
of the optical spectral weight. 
Using the abbreviations  $c_\theta$\,=\,$\cos\theta_d$ and $s_\theta$\,=\,$\sin\theta_d$,  
we find the matrix elements 
\begin{eqnarray}
\nonumber
|M_{d_{\pm 1}}(^4\!A_2)|^2  & = & \left[ \frac{2}{3} \beta^2 \, c^4_\theta  +  2 \, c^2_\theta s^2_\theta 
+ \frac{1 + 2 \beta^2}{3}   \, s^4_\theta \right] 
\frac{t^4_{pd\pi}}{\Delta^2_{\rm CT}}
\end{eqnarray}
\begin{eqnarray}
\nonumber
|M_{d_{\pm 1}}(^2\!T_1)|^2 & = & \left[c^4_\theta + \beta^2 \, c^2_\theta s^2_\theta 
 +  \frac{5 + 4 \beta^2}{8}  \,  s^4_\theta \right]  
\frac{t^4_{pd\pi}}{\Delta^2_{\rm CT}}
\end{eqnarray}
\begin{eqnarray}
\nonumber
|M_{d_{\pm 1}}(^2E)|^2   & = & \left[\frac{4}{3} \beta^2 \, c^4_\theta  + c^2_\theta s^2_\theta  
 + \frac{1 + 2 \beta^2}{6}   \, s^4_\theta \right]  
\frac{t^4_{pd\pi}}{\Delta^2_{\rm CT}}  \, .
\end{eqnarray}
The corresponding spectral weights are plotted as a function of $\Delta_{\rm CF}/\zeta$  for $\beta$\,=\,1/3 in Fig.\ \ref{fig:SWratio}, 
which also shows the behavior of $\cos\theta_d$. 
The spectral weight of the $^4A_2$ multiplet (red), which corresponds to peak A, is boosted by spin-orbit coupling. This directly reflects the dependence of the ground state wavefunction on $\Delta_{\rm CF}/\zeta$. 
A decrease of $\Delta_{\rm CF}/\zeta$ yields an enhanced admixture of $|xz^2,0\rangle$ to $|d_{\pm 1}\rangle$, 
see Eqs.\ (\ref{eq:doup1}) and (\ref{eq:doum1}). The doubly occupied $xz$ orbital allows for contributions 
to peak A via diagonal hopping processes from $xz_i$ to $xz_j$ or from $xy_i$ to $xy_j$. 
In the spectral weight, their prefactor $\alpha^2$\,=\,1 is larger than $\beta^2$\,=\,1/9 for the 
off-diagonal contributions, see Eq.\ (\ref{eq:tmat}).
In contrast, the spectral weights of the $^2T_1$ and $^2E$ multiplets do not vary strongly for 
$\Delta_{\rm CF}/\zeta \gtrsim 1.5$.

\begin{figure}[bt]
	\includegraphics[width=0.9\columnwidth,clip]{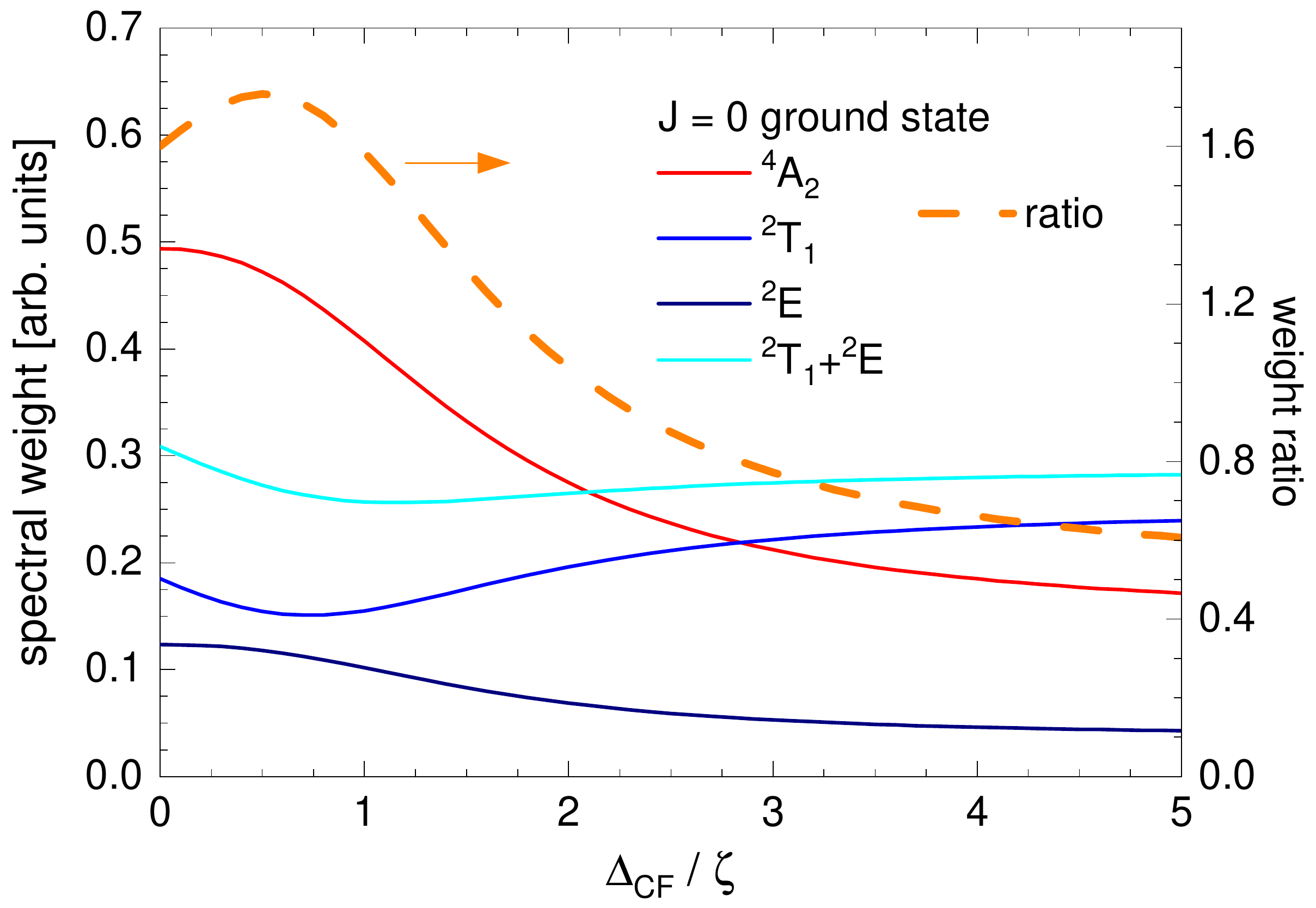}
	\caption{Results for the singlet ground state $|s\rangle_i |s\rangle_j$.  
		Left axis: Spectral weight of peak A ($^4A_2$ multiplet, red) and of the 
		two contributions to peak B ($^2T_1$ and $^2E$, blue). 
		Right axis (dashed orange line): The spectral weight ratio $SW({}^4\!A_2)/SW(^2T_1 \! + \! {} ^2\!E)$ 
		is much larger than for the doublet ground state considered in Fig.\ \ref{fig:SWratio}. 
	}
	\label{fig:SWratioSinglet}
\end{figure}

Our calculations reveal a strong dependence of the ratio $SW({}^4A_2)/SW({}^2T_1\! + \!{}^2\!E)$ on $\Delta_{\rm CF}/\zeta$ (dashed orange),  
which allows us to estimate this important parameter. 
The shaded area in Fig.\ \ref{fig:SWratio} indicates the range in which the calculation agrees with 
the experimental result $SW_{\rm A}/SW_{\rm B}$\,=\,1/3 to 1/4. 
For the chosen hypothetical ground state $|d_{+1}\rangle_i |d_{-1}\rangle_j$  
this yields $\Delta_{\rm CF}/\zeta$\,=\,$2.2 \pm 0.3$, which agrees with a recent analysis 
of RIXS data \cite{Gretarsson19}. 
However, our result denotes a lower limit of $\Delta_{\rm CF}/\zeta$ since it neglects contributions 
of the singlet. 
We find that adding a finite singlet weight to $|d_{+1}\rangle_i |d_{-1}\rangle_j$ raises the result for $\Delta_{\rm CF}/\zeta$. 
To illustrate this, we plot the spectral weights for the two-site singlet ground state $|s\rangle_i |s\rangle_j$
in Fig.~\ref{fig:SWratioSinglet}. For $\Delta_{\rm CF}/\zeta \gtrsim 1$, the spectral weight of $^4A_2$ is strongly enhanced 
while that of $^2T_1\! + \!{} ^2E$ is strongly reduced compared to the doublet-type ground state. 
The much larger spectral weight ratio exceeds the experimental value 1/3 to 1/4 for any $\Delta_{\rm CF}/\zeta$.

\begin{figure}[bt]
	\includegraphics[width=0.9\columnwidth,clip]{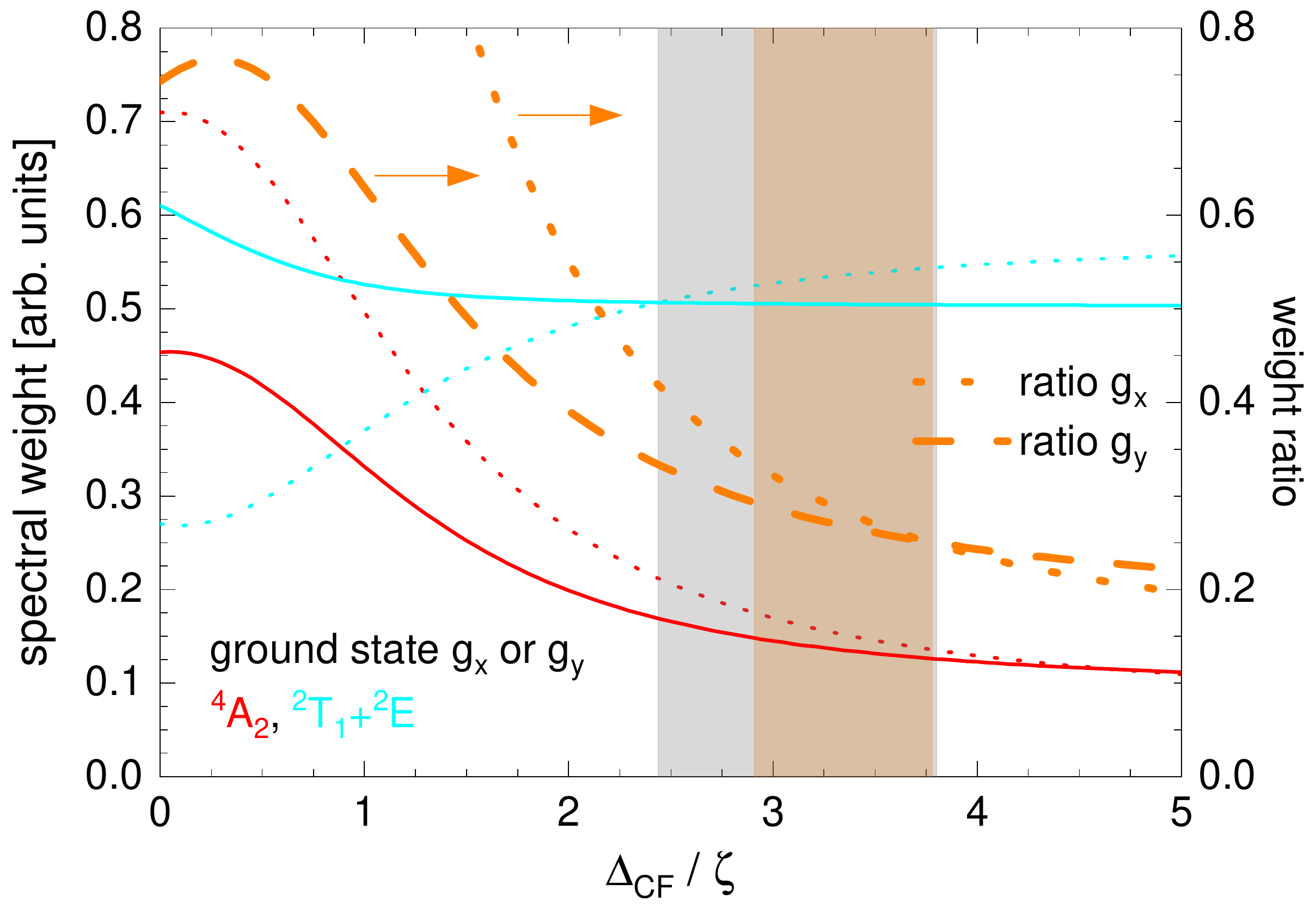}
	\caption{Spectral weights for the realistic ground states $|g_x\rangle$ (dotted) and $|g_y\rangle$ (solid lines) for $\sin^2(\Phi)$\,=\,1/2, see Eq.\ (\ref{eq:gx}).
	Data for $^4\!A_2$ (red) and $^2T_1$\,$+$\,$^2\!E$ (cyan) correspond to peaks A and B, respectively. 
	Orange curves depict their ratio $SWR$ (right axis),  
	and the shaded range gives the values of $\Delta_{\rm CF}/\zeta$ for which 
	1/4\,$\leq SWR \leq$\,1/3 (orange for $|g_x\rangle$, grey for $|g_y\rangle$; 
	cf.\ Fig.\ \ref{fig:singlet_weight}).
	}
	\label{fig:gxgyratio}
\end{figure}

\begin{figure}[bt]
	\includegraphics[width=0.9\columnwidth,clip]{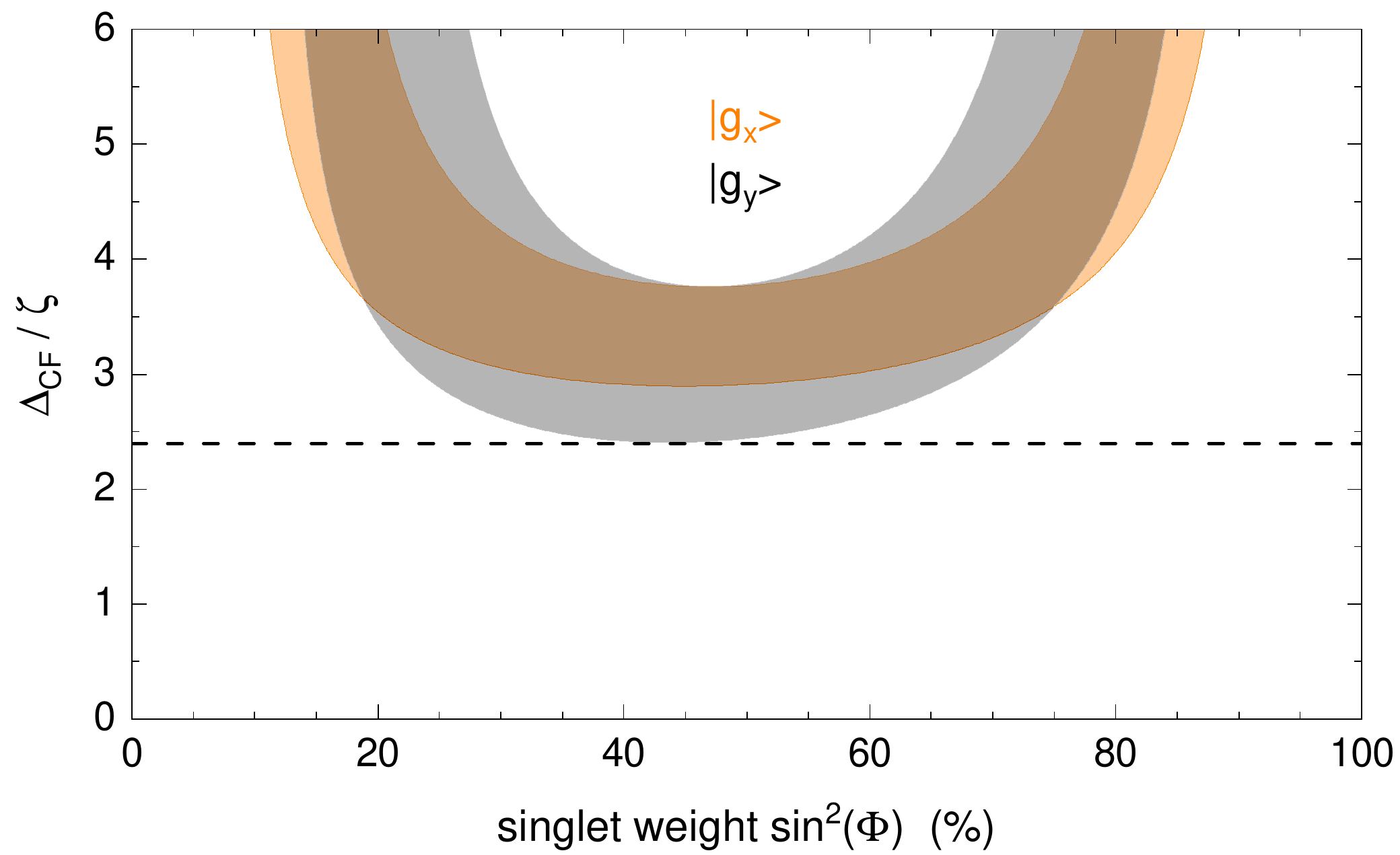}
	\caption{Results for the realistic ground states $|g_x\rangle$ (orange, Eq.\ \ref{eq:gx}), and $|g_y\rangle$ (grey). 
	The shaded range depicts the values of $\Delta_{\rm CF}/\zeta$ for which the 
	calculated spectral weight ratio $SW({}^4\!A_2)/SW(^2T_1+{} ^2\!E)$ lies within 
	the experimental result 1/3 to 1/4. 
	The dashed line highlights the lower limit $\Delta_{\rm CF}/\zeta \geq 2.4$.
	}
	\label{fig:singlet_weight}
\end{figure}

For a more realistic scenario, we have to take into account that the moments 
	in Ca$_2$RuO$_4$ are lying within the $ab$ plane in the ordered phase. 
	This requires to consider the superposition of $|d_x\rangle$ (or equivalently $|d_y\rangle$) 
	with the singlet \cite{Akbari14,Feldmaier20,Strobel21},
	\begin{eqnarray}
		\nonumber
		|g_x \rangle &\! = \! &  \big( \, \sin \Phi \, |s\rangle_i  +  \cos \Phi \, |d_x \rangle_i \, \big) \\
		\label{eq:gx}
		& \cdot &                \big( \, \sin \Phi \, |s\rangle_j  -  \cos \Phi \, |d_x \rangle_j \, \big)  
	\end{eqnarray}
	where $|s \rangle$ and $|d_x\rangle$ depend on $\Delta_{\rm CF}/\zeta$, 
	see Eqs.\ (\ref{eq:sing}) -- (\ref{eq:doum1}) and (\ref{eq:dxy}),
	and $0 \leq \Phi \leq \pi$. 
	The $+/-$ signs denote the relative phases between $\sin \Phi \, |s\rangle$ and 
	$\cos \Phi \, |d_x \rangle$ for which we obtain the lower bound of $\Delta_{\rm CF}/\zeta$ for this type of ground state. 
For each $\Phi$ we calculate the spectral weights. As an example, Fig.~\ref{fig:gxgyratio} 
plots the curves for equal weights of singlet and doublet, $\sin^2(\Phi)$\,=\,1/2. This yields the range of $\Delta_{\rm CF}/\zeta$ in which the calculated spectral weight ratio agrees with 
the experimental result 1/3 to 1/4 at 15\,K, see orange shaded range in Figs.\ \ref{fig:gxgyratio} and \ref{fig:singlet_weight}. 
The grey shaded range shows the corresponding solution for $|g_y\rangle$ which differs from 
the result for $|g_x\rangle$ since we consider a bond parallel to $x$. 
Experimentally, we could not resolve any in-plane anisotropy, hence our data correspond 
to an average over both cases. 
For this realistic ground state, we find a lower bound $\Delta_{\rm CF}/\zeta \geq 2.4$ that is reached for singlet weight $\sin^2(\Phi)$ in the vicinity of 1/2. 
Note that a variational cluster approach indeed points towards similar weights of singlet and doublet for parameters applicable to the case of Ca$_2$RuO$_4$ \cite{Feldmaier20}.

To estimate the upper bound of $\Delta_{\rm CF}/\zeta$, we use 
the RIXS peak frequency 0.34\,eV \cite{Das18,Gretarsson19} as the upper limit of $\Delta_{\rm CF}$, 
see Fig.\ \ref{fig:energiesLS}. 
For $\zeta$, different values were claimed, e.g., 0.11\,eV \cite{Zhang20}, 0.13\,eV \cite{Gretarsson19}, or 0.4\,eV \cite{Fatuzzo15}. 
As a lower limit, we use $\zeta$\,=\,0.08\,eV which was found in 
an analysis of inelastic neutron data \cite{Sarte20} and was also used to analyze RIXS data of 
Ca$_2$RuO$_4$ \cite{Das18}. This yields $\Delta_{\rm CF}/\zeta \lesssim 4$. 
Altogether, our final result hence reads $2.4 \leq \Delta_{\rm CF}/\zeta \lesssim 4$. 
In this parameter range, we deduce 0.31\,eV\,$\lesssim \Delta_{\rm CF} \lesssim$\,0.34\,eV 
by identifying the RIXS peak at 0.34\,eV \cite{Das18,Gretarsson19} with the average 
excitation energy plotted in the right panel of Fig.\ \ref{fig:energiesLS}, which finally 
yields 0.08\,eV\,$\lesssim \zeta \lesssim 0.14$\,eV.\@ 
Our findings agree very well with LDA+DMFT calculations that find $\Delta_{\rm CF}$\,=\,0.32\,eV, 
$\zeta$\,=\,0.106\,eV, and hence $\Delta_{\rm CF}/\zeta$\,$\approx$\,3. 	
According to different theoretical approaches, 
these parameters are safely in the range of AF order with moments lying in the $ab$ plane 
\cite{Mohapatra20,Strobel21}.

\subsubsection{Temperature dependence}

In Ca$_2$RuO$_4$, also the behavior at finite temperature is characterized by the competition between spin-orbit coupling and crystal field \cite{Lotze21}. 
In general, the temperature dependence of the spectral weight reflects the behavior of nearest-neighbor spin and orbital correlations. Due to the prominent role of $\zeta$, 
these need to be addressed simultaneously in Ca$_2$RuO$_4$.  
Without more detailed knowledge on the precise $T$\,=\,0 ground state on the lattice, 
we do not attempt to quantitatively address the temperature dependence for finite $\zeta$. 
Qualitatively, however, the experimental data can be explained by the pronounced 
temperature dependence of $\Delta_{\rm CF}$.  
With increasing temperature, the decrease of $\Delta_{\rm CF}/\zeta$ yields a pronounced increase of peak A.\@ This occurs in all the examples that we studied, 
see Figs.\ \ref{fig:SWratio} -- \ref{fig:gxgyratio}. 
For peak B, our experimental data show an increase of spectral weight with increasing temperature, i.e., decreasing $\Delta_{\rm CF}/\zeta$. This trend is 
reproduced for the realistic ground state $|g_y\rangle$, see 
Fig.\ \ref{fig:gxgyratio}.

\section{Conclusion}

In the layered ruthenates Ca$_{2-x}$Sr$_x$RuO$_4$, the competition of spin-orbit coupling $\zeta$\,=\,$2\lambda$ 
and crystal-field splitting $\Delta_{\rm CF}$ has been discussed controversially over many years. 
We have shown that the optical spectral weight of Mott-Hubbard excitations in $4d^4$ compounds is a direct measure 
of the central parameter $\Delta_{\rm CF}/\zeta$. 
The sensitivity of the optical data is based on the pronounced effect that $\Delta_{\rm CF}/\zeta$ 
has on the ground state wavefunction. The optical matrix elements reflect the orbital occupation 
which strongly depends on $\Delta_{\rm CF}/\zeta$. 
This is different from the limiting cases $\Delta_{\rm CF}$\,=\,0 or $\zeta$\,=\,0, 
where the ground state is insensitive to the size of the finite one of the two parameters, 
at least over large ranges. 
In $3d$ transition-metal compounds with negligible spin-orbit coupling, the spectral weight of 
Mott-Hubbard excitations reflects nearest-neighbor spin and orbital correlations but it does not 
directly allow to determine the size of the crystal field. This rather requires to consider the 
excitation energies. The possibility to determine $\Delta_{\rm CF}/\zeta$ directly from the spectral weight 
in compounds with sizable spin-orbit coupling such as the $4d^4$ ruthenates is the central result 
of this study.

In Ca$_2$Ru$_{0.99}$Ti$_{0.01}$O$_4$, we have focused on the lowest Mott-Hubbard excitations 
at about 1\,eV and 2\,eV in the optical conductivity. 
Based on the excitation energies, we observe a pronounced temperature dependence of $\Delta_{\rm CF}$, 
as expected from the behavior of the octahedral distortion. 
In particular we find that $\Delta_{\rm CF}$ is strongly suppressed at room temperature. 
At low temperature, the optical spectral weight allows us to estimate 
a lower bound $\Delta_{\rm CF}/\zeta \geq 2.4$ which is reached for a ground state with roughly similar weights of singlet and doublet. Altogether we find 
0.31\,eV\,$\lesssim \Delta_{\rm CF} \lesssim$\,0.34\,eV and 
0.08\,eV\,$\lesssim \zeta \lesssim 0.14$\,eV.\@

On the one hand, one may conclude that Ca$_2$RuO$_4$ is firmly rooted in the range of dominant crystal field. For instance the energy scales of the lowest excitations are not given by $\zeta$ but rather by $(\zeta/2)^2/\Delta_{\rm CF}$ and $\Delta_{\rm CF}$. 
On the other hand, spin-orbit coupling still plays a most important role in Ca$_2$RuO$_4$. 
For the low-energy magnetic excitations, spin-orbit coupling may be treated 
perturbatively but it gives rise to an unusually strong anisotropy. 
Considering electronic excitations such as predominantly local crystal-field-type excitations or intersite Mott-Hubbard excitations, the matrix elements are strongly affected by spin-orbit coupling and the local non-magnetic singlet ground state poses a challenge for quantitative calculations. 
In the end, the captivating character of Ca$_2$RuO$_4$ is based on the \textit{competition} of spin-orbit coupling and crystal field rather than on the dominance of one of them.

\section*{Acknowledgement}
We acknowledge funding from the Deutsche Forschungsgemeinschaft (DFG, German Research Foundation) – Project
No. 277146847 – CRC 1238 (Projects A02, B02, and C02). 
M.H. is supported by the Knut and Alice Wallenberg Foundation as part of the Wallenberg Academy Fellows project.

\section*{Appendix}

\subsection{Anisotropy}

Our analysis focuses on peaks A and B, i.e., Mott-Hubbard excitations that do not contribute to the $c$-axis response of Ca$_2$RuO$_4$ \cite{Jung03}.  
Their energies provide the basis for Fig.\ \ref{fig:Delta}, and the spectral weight ratio $SW_A/SW_B$ at 15\,K yields our estimate of $\Delta_{\rm CF}/\zeta$. In the chosen measurement geometry, the pseudo-dielectric function allows us to reliably determine these quantities \cite{Aspnes80}. This is supported by the good overall agreement of our experimental data with reflectivity-based results \cite{Jung03,Lee02}. Moreover, this claim is corroborated by an alternative approach which explicitly considers the anisotropy. In this case, we simultaneously analyze our ellipsometric data and 
$\sigma_{\parallel c}$, the optical conductivity for polarization parallel to $c$ 
reported by Jung \textit{et al.} \cite{Jung03}. This allows us to determine $\sigma_{\perp c}$. 
Note that the $c$-axis data shows little temperature dependence.

\begin{figure}[t]
	\includegraphics[width=\columnwidth,clip]{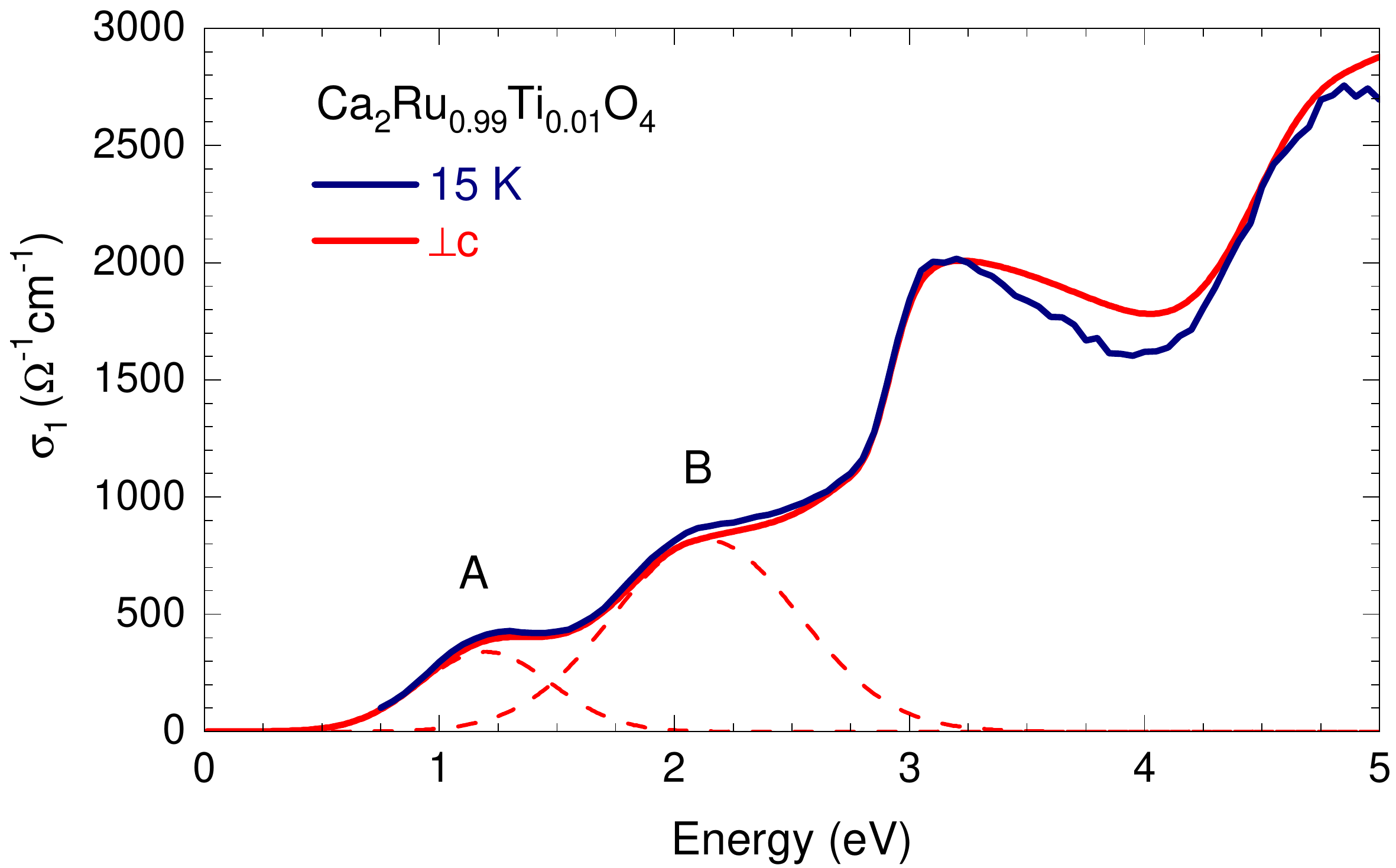}
	\caption{Taking into account the $c$-axis response \cite{Jung03} in a uniaxial model (red) yields a very similar result as the analysis based on the 
	pseudo-dielectric function (blue), cf.\ Fig.\ \ref{fig:sig1}a.
	This applies in particular to peaks A and B (dashed). 
	}
	\label{fig:aniso}
\end{figure}

In Fig.\ \ref{fig:aniso} we compare $\sigma_{\perp c}$ at 15\,K with the data plotted in Fig.\ \ref{fig:sig1}a. The two curves are very similar. For peaks A and B, we find no effect on the peak energies and minor changes of the spectral weights.\@ For the spectral weight ratio $SW_A/SW_B$ we find 0.29, in perfect agreement with the values 0.30 and 0.27 discussed in the main text, see Fig.\ \ref{fig:fit}.

\subsection{Optical matrix elements}
\label{AppA}

\begin{table}[b]
	\begin{tabular}{cccc}
		$c^\dagger_{xz.\sigma}$ & $|xy^2, 1\rangle$ & $|xy^2,0\rangle$ &  $|xy^2,-1\rangle$  \\ 
		\hline 
		$\langle t_{2g}^5 (^2T_2,yz\uparrow) |$	 & 1 ($\downarrow$) & $-1/\sqrt{2}$ ($\uparrow$) &  0 \\
		$\langle t_{2g}^5 (^2T_2,yz\downarrow) |$  & 0              &  $1/\sqrt{2}$ ($\downarrow$) &  -1 ($\uparrow$)\\ 
		& & & \\
		$c_{xz,\sigma}$        &  $|xy^2, 1\rangle$ & $|xy^2,0\rangle$ &  $|xy^2,-1\rangle$  \\ 
		\hline 
		$\langle t_{2g}^3(^2T_1, yz\uparrow) |$   & $-1/\sqrt{2}$ ($\uparrow$) &  $-1/2$ ($\downarrow$) &  0 \\ 
		$\langle t_{2g}^3(^2T_1, yz\downarrow) |$ & 0         & $-1/2$ ($\uparrow$)   &  $-1/\sqrt{2}$ ($\downarrow$) \\
		& & & \\
		$c_{xy,\sigma}$        &  $|xy^2, 1\rangle$ & $|xy^2,0\rangle$ &  $|xy^2,-1\rangle$  \\ 
		\hline 
		$\langle t_{2g}^3(^4\!A_2, 3/2) |$  & -1 ($\downarrow$)   & 0   &  0 \\
		$\langle t_{2g}^3(^4\!A_2, -3/2) |$ & 0                   & 0   &  1 ($\uparrow$) \\
		$\langle t_{2g}^3(^4\!A_2, 1/2) |$  & $1/\sqrt{3}$ ($\uparrow$) & $-\sqrt{2/3}$ ($\downarrow$)   &  0 \\
		$\langle t_{2g}^3(^4\!A_2, -1/2) |$ & 0                   &  $\sqrt{2/3}$ ($\uparrow$) &  $-1/\sqrt{3}$ ($\downarrow$) \\
		$\langle t_{2g}^3(^2E, a\uparrow) |$   & $-\sqrt{2/3}$ ($\uparrow$) & $-1/\sqrt{3}$ ($\downarrow$)   &  0 \\
		$\langle t_{2g}^3(^2E, a\downarrow) |$ & 0             &  $1/\sqrt{3}$ ($\uparrow$) & $\sqrt{2/3}$ ($\downarrow$)
	\end{tabular} 
	\caption{From top to bottom: Single-site matrix elements 
		$\langle d^5_{m^\prime} | c^\dagger_{xz,\sigma} |d^4_{n^\prime} \rangle $, 
		$\langle d^3_m | c_{xz,\sigma} |d^4_n \rangle $, and 
		$\langle d^3_m | c_{xy,\sigma}|d^4_n \rangle $ (bottom) 	
		for double occupancy of the $xy$ orbital in the $d^4$ initial state. 
		The allowed value of spin $\sigma$ is given in brackets for each finite entry.  }
	\label{tab:singlesite}
\end{table}

To calculate the contribution of Mott-Hubbard excitations to the optical conductivity $\sigma_1(\omega)$, 
we consider hopping between nearest-neighbor Ru sites $i$ and $j$ at a distance $d_{\rm Ru}$ and polarization 
parallel to this Ru-Ru bond, 
\begin{equation}
\sigma_1(\omega) = \frac{d_{\rm Ru}^2}{N_{\rm conf}} \!
\sum_{\substack{m,m^\prime,\\ n,n^\prime}} \frac{|M_{n,n^\prime}^{m,m^\prime}|^2}{E_{\rm MH,m}}\, \, \delta(\hbar\omega-E_{\rm MH,m}) 
\end{equation}
where $m$ and $m^\prime$ denote the different $d^3$ and $d^5$ states, respectively, 
$n$ and $n^\prime$ correspond to the possible $d^4$ states in the ground state, 
$M_{n,n^\prime}^{m,m^\prime}$ is the matrix element, 
$E_{\rm MH,m}$ the energy as given in Eq.\ (\ref{eq:energyMH}), 
and $N_{\rm conf}$ denotes the number of possible configurations in the ground state. 
At $T$\,=\,$0$, the N\'eel state shows finite sublattice magnetization, and one finds $N_{\rm conf}$\,=\,2. 
For $\zeta$\,=\,0 and double occupancy of the $xy$ orbital, the two configurations are 
$|xy^2, S_z\rangle_i |xy^2, -S_z\rangle_j$ for $S_z$\,=\,$\pm 1$. 
In this case, the $S_z$\,=\,0 state corresponds to a magnon excitation. 
For finite $\zeta$, the doublet ground state $|d_{J_z}\rangle_i |d_{-J_z}\rangle_j$ with $J_z$\,=\,$\pm 1$
also shows $N_{\rm conf}$\,=\,2. 
A magnetically fully disordered state for $\zeta$\,=\,0 and $\Delta_{\rm CF} > 0$ shows $N_{\rm conf}$\,=\,9. 
The matrix elements are given by 
\begin{eqnarray}
\nonumber
& & |M_{n,n^\prime}^{m,m^\prime}|^2 \\
\nonumber & = & \sum_{k,k^\prime} \left| \langle d_{k,m}^3 d_{k^\prime,m^\prime}^5| \!\!
\sum_{\tau,\tau^\prime,\sigma} \!\!
t_{\tau\tau^\prime} c^\dagger_{k^\prime,\tau^\prime\sigma} c_{k,\tau\sigma} 
| d_{k,n}^4 d_{k^\prime,n^\prime}^4 \rangle  \right|^2 
\\ \nonumber
& = &   \sum_{k,k^\prime}\left|- \sum_{\tau,\tau^\prime,\sigma}  t_{\tau\tau^\prime} 
\langle d_{m}^3|                c_{\tau\sigma}  | d_{n}^4\rangle_k
\langle d_{m^\prime}^5|  c^\dagger_{\tau^\prime\sigma}   | d_{n^\prime}^4 \rangle_{k^\prime} \right|^2 \, .
\end{eqnarray}
where $t_{\tau\tau^\prime}$ is the hopping matrix element between orbitals $\tau$ and $\tau^\prime$ 
on adjacent sites, $\sigma \in \{\uparrow,\downarrow\}$ refers to the electron spin, and 
$kk^\prime \in \{ij, ji \}$. 
The restriction to nearest-neighbor sites is justified because of the small hopping matrix elements 
between further neighbors.
The last line uses single-site matrix elements which facilitates the calculation. 
For double occupancy of the $xy$ orbital in the ground state, the relevant single-site matrix elements 
are given in Tab.\ \ref{tab:singlesite}.

\vspace*{1cm}
\subsection{Cubic multiplets}

Here we give the cubic multiplets that appear in Tab.\ \ref{tab:singlesite} \cite{Zhang17}.  

\begin{align}
|t_{2g}^5 (^2T_2,yz \sigma) \rangle  & =  c^\dagger_{yz \sigma} c^\dagger_{xz \uparrow} c^\dagger_{xz \downarrow} c^\dagger_{xy \uparrow} c^\dagger_{xy \downarrow} |0\rangle 
\\
|t_{2g}^3 (^4\!A_2, 3\sigma)\rangle & =  c^\dagger_{xz\sigma} c^\dagger_{yz\sigma} c^\dagger_{xy\sigma} |0\rangle 
\label{eq:4A232up}\\
\nonumber
|t_{2g}^3 (^4\!A_2, \sigma)\rangle & =  \frac{1}{\sqrt{3}} \left(
  c^\dagger_{xz\sigma} c^\dagger_{yz\sigma} c^\dagger_{xy-\sigma}
+ c^\dagger_{xz-\sigma} c^\dagger_{yz\sigma} c^\dagger_{xy\sigma} \right. \\
 &  \left. 
 + c^\dagger_{xz\sigma} c^\dagger_{yz-\sigma} c^\dagger_{xy\sigma}
\right) |0\rangle \, .
\label{eq:4A212up}
\\
\nonumber
|t_{2g}^3 (^2E, a,\sigma)\rangle & =  \frac{1}{\sqrt{6}} \left(-2 \, c^\dagger_{xz\sigma} c^\dagger_{yz\sigma} c^\dagger_{xy-\sigma}
+  c^\dagger_{xz-\sigma} c^\dagger_{yz\sigma} c^\dagger_{xy\sigma} \right. 
\\ \label{eq:2Ea} 
&  \left. + c^\dagger_{xz\sigma} c^\dagger_{yz-\sigma} c^\dagger_{xy\sigma}
\right) |0\rangle 
\\
|t_{2g}^3 (^2T_{1}, yz \,\sigma)\rangle & =  \frac{1}{\sqrt{2}} 
 \left[ c^\dagger_{yz\sigma} c^\dagger_{xy\uparrow} c^\dagger_{xy\downarrow} 
- c^\dagger_{xz\uparrow} c^\dagger_{xz\downarrow} c^\dagger_{yz\sigma}\right] |0\rangle 
\end{align}


\begin{thebibliography}{99}
\bibitem{CommentJH}	
For the $t_{2g}$ shell one finds $U_{t_{2g}}$\,=\,$F^0+(4/49)(F^2+F^4)$ and 
$J_{H,t_{2g}}$\,=\,$(3/49)F^2 +(20/441)F^4$ \cite{Georges13}, where $F^0$, $F^2$, and $F^4$ 
are Slater parameters. 
As common, we use parameters referring to the entire $d$ shell, with $U$\,=\,$F^0$ and 
$J_H \approx J_{H,t_{2g}}/0.77$ \cite{Georges13}. Both approaches nominally yield the 
same expressions for the Mott-Hubbard excitation energies given in Tab.\ \ref{tab:multipletenergies}.
\end{thebibliography}
\end{document}